\documentclass[aps,superscriptaddress,prl,reprint,secnumarabic]{revtex4-2}
\usepackage{soul}
\usepackage{dsfont}
\usepackage[utf8]{inputenc}
\usepackage{physics}
\usepackage{float}
\usepackage{dsfont}
\usepackage{appendix}
\usepackage{xcolor}
\usepackage{graphicx,subfigure,bbm}
\usepackage{ amssymb }
\usepackage[normalem]{ulem}
\usepackage{bigints}

\usepackage{tabularray}
\usepackage{hyperref}
\hypersetup{
	colorlinks,
	linkcolor={red!50!black},
	citecolor={blue!50!black},
	urlcolor={blue!80!black}
}
\usepackage{mathtools}
\usepackage{wasysym}
\usepackage{chngcntr}
\counterwithout{table}{section}
\usepackage{url}
\usepackage{caption}
\usepackage{subcaption}

\usepackage{ragged2e}

\DeclareCaptionJustification{justified}{\justifying}
\def \beq {\begin{equation}}
\def \edq {\end{equation}}
\def \bes {\begin{subequations}}
\def \eds {\end{subequations}}
\def \beqn {\begin{equation*}}
\def \edqn {\end{equation*}}

\def \dag {\dagger}

\def \calg {{\cal{G}}}

\definecolor{crimson}{RGB}{255,102,255}

\newcommand{\me}{\text{ME}}

\newcommand{\wc}{\text{WC}}

\begin{document}
\title{
%Quantum Kinetic Uncertainty Relations for Coherent Transport
Quantum Kinetic Uncertainty Relations in Mesoscopic Conductors at Strong Coupling
}

\author{Gianmichele Blasi}
	\affiliation{Department of Applied Physics, University of Geneva, 1211 Geneva, Switzerland}
    \affiliation{Instituto de Física Interdisciplinar y Sistemas Complejos IFISC (CSIC-UIB), E-07122 Palma de Mallorca, Spain}
    % \email{gianmichele.blasi@unige.ch}
\author{Ricard Ravell Rodríguez}
 \affiliation{Instituto de Física Interdisciplinar y Sistemas Complejos IFISC (CSIC-UIB), E-07122 Palma de Mallorca, Spain}
 \affiliation{ICFO-Institut de Ciències Fotòniques, The Barcelona Institute of
Science and Technology, 08860 Castelldefels (Barcelona), Spain}
\author{Mykhailo Moskalets}
 \affiliation{Instituto de Física Interdisciplinar y Sistemas Complejos IFISC (CSIC-UIB), E-07122 Palma de Mallorca, Spain}
\affiliation{Department of Metal and Semiconductor Physics\char`,~NTU “Kharkiv Polytechnic Institute”, 61002 Kharkiv, Ukraine} 
 \author{Rosa López}
 \affiliation{Instituto de Física Interdisciplinar y Sistemas Complejos IFISC (CSIC-UIB), E-07122 Palma de Mallorca, Spain}
 \author{G\'eraldine Haack}
	\affiliation{Department of Applied Physics, University of Geneva, 1211 Geneva, Switzerland}

\begin{abstract}
Kinetic Uncertainty Relations (KURs) set fundamental limits on the precision of nonequilibrium transport by bounding the signal-to-noise ratio of currents in terms of the dynamical activity, a quantity that counts exchange events between a system and its reservoirs. This framework is well established in the weak-coupling regime, where transport occurs via well-defined, particle-like tunneling processes. 
At strong coupling, however,
quantum coherence challenges both the validity of standard KURs and the notion of activity itself.
In this Letter, we introduce a generalized definition of dynamical activity valid at arbitrary
system-reservoir coupling, and show that it leads to a breakdown of standard KURs at strong
coupling. Building on this result, we derive and prove a novel uncertainty relation, denoted Quantum KUR
(QKUR), which provides a genuine quantum extension of KUR, accounting for intrinsic quantum coherent contributions of the generalized activity.  
We demonstrate that the generalized activity
reduces to the standard master-equation definition in the weak-coupling regime for generic
systems of $N$ coupled quantum dots described by a quadratic Hamiltonian, and analyze the resulting QKUR bound in paradigmatic quantum-coherent
mesoscopic devices, including single- and double-quantum dot systems and a quantum point contact.

\end{abstract}

% \date{\today}
\maketitle
%%On fluctuation theorems and bounds on SNR

\emph{Introduction} - A central theme in non-equilibrium and stochastic thermodynamics is the study of fluctuation theorems~\cite{Jarzynski1997,Crooks1999,Jarzynski2011,Seifert2012, Seifert2019, Dechant2020, Aslyamov2025}. In this context, uncertainty relations have been derived to place fundamental constraints on current fluctuations. They often take the form of a bound on the signal-to-noise ratio (SNR), $I^2_\alpha/S_{\alpha \alpha} \leq \xi$, with $I_{\alpha}$ the average current (of particles, charge, or energy) in reservoir $\alpha$, and $S_{\alpha\alpha}$ its variance, corresponding to the zero-frequency component of the current autocorrelation function. The quantity $\xi$ is an upper bound that depends on the context. 

For Thermodynamic Uncertainty Relations (TURs), this bound is given by the entropy production rate $\xi_{\text{TUR}} = \sigma/2k_B$ (with $k_B$ the Boltzmann constant)~\cite{Barato2015,Gingrich2016,Pietzonka2018, Hasegawa2019, Timpanaro2019,  Hasegawa2019b, Horowitz2020, Falasco2020, Ray2023, Ptaszyski2024}. At their core, TURs unveil a precision-energy trade-off corresponding to the energy cost (dissipation) required to achieve higher precision in measurements. First derived within the framework of classical stochastic thermodynamics, TURs have been investigated in a variety of quantum systems, including periodically driven and measured systems~\cite{Vo2020, Koyuk2020,Miller2021,Hasegawa2023,Yunoki2025, Potts2019} and hybrid superconducting devices~\cite{Lopez2023,Taddei2023,Manzano2023}. Violations have been demonstrated in quantum coherent setups~\cite{Brandner2018, Agarwalla2018, Ptaszyski2018, Liu2019, Cangemi2020, Liu2020, Kalaee2021, Menczel2021, Ehrlich2021, Timpanaro2021, Gerry2022}, motivating the derivation of looser bounds inspired by quantum information theory~\cite {Hasegawa2022, Guarnieri2019} and a novel bound valid in the quantum regime for arbitrary non-equilibrium conditions~\cite{Brandner2025}.

\begin{figure}[t!]
{{\includegraphics[width=.45\textwidth]{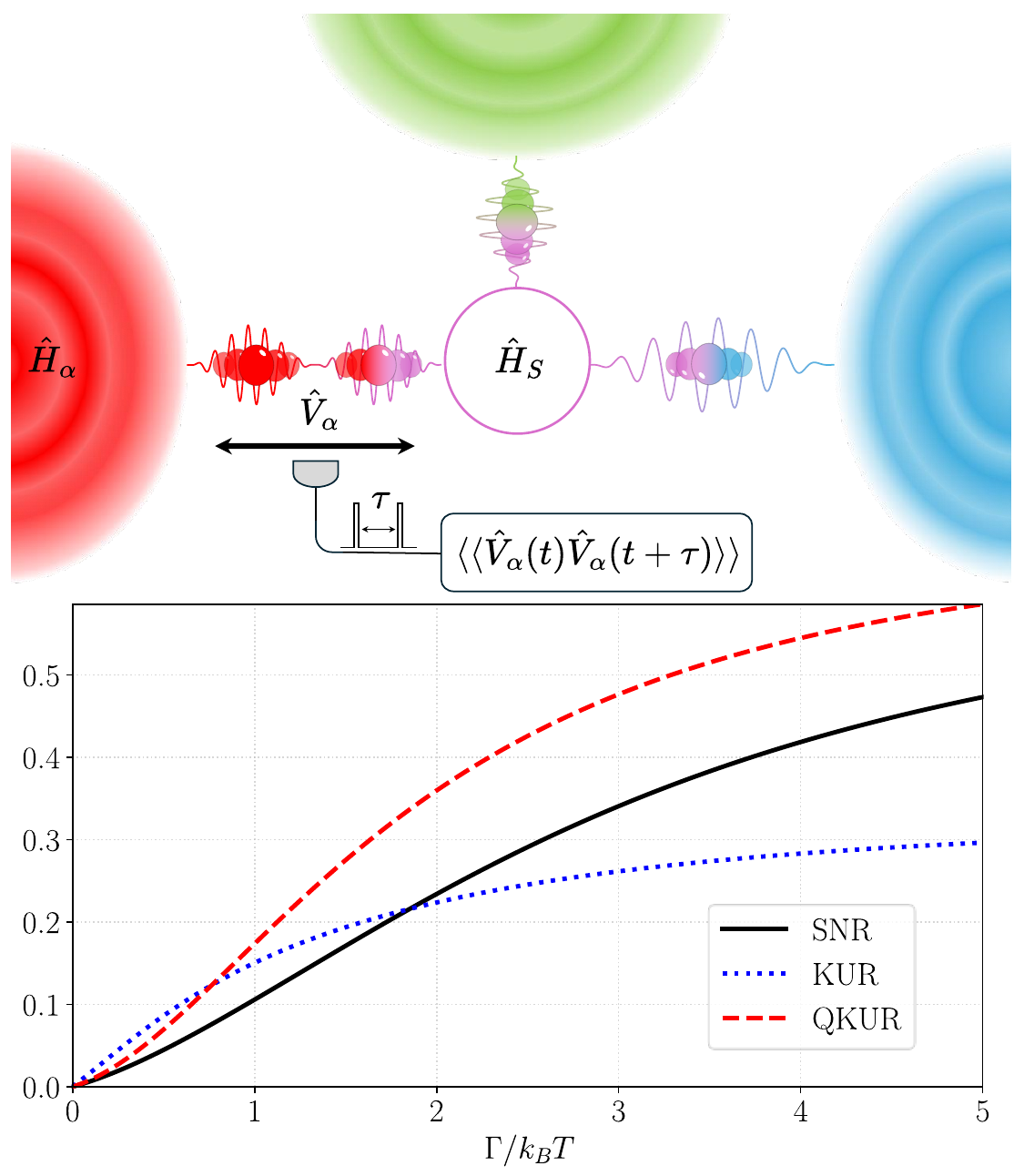} }}
\caption{\textit{Top panel} -- Scheme of a generic multiterminal setup. A quantum system ($\hat{H}_S$) couples to reservoirs ($\hat{H}_\alpha$) via interaction Hamiltonians ($\hat{V}_\alpha$). The generalized dynamical activity is defined through two-time exchange-rate fluctuations $\langle\langle \hat{V}_\alpha(t)\hat{V}_\alpha(t+\tau)\rangle\rangle$.
\textit{Bottom panel} — SNR, \textit{i.e.} $I^2_L/S_{LL}=I^2_R/S_{RR}$ (solid black), for a two-terminal SQD with equal-temperature reservoirs, as a function of coupling strength $\Gamma/k_B T$, characterized by the transmission probability $\mathcal{T}_{RL}(\epsilon)=\Gamma_L\Gamma_R/\left[\Gamma^2/4+(\epsilon-\epsilon_d)^2\right]$~\cite{Blasi2024}. The KUR bound $\xi_{\rm KUR}$ (blue dotted) and the QKUR bound $\xi_{\rm QKUR}$ (red dashed), calculated from Eq.~\eqref{eq:QKUR}, are shown. KUR is valid only at small $\Gamma$, whereas QKUR holds for any coupling. Parameters: $\epsilon_d/k_BT=3$, $\Delta\mu/k_BT=2$.}
\label{fig1}
\end{figure}

Another family of uncertainty relations that has attracted increasing interest in recent years is the Kinetic Uncertainty Relations (KURs). Unlike entropy-based bounds, the bound $\xi_{\text{KUR}} = \mathcal{A}$ in KURs is based on the dynamical activity (or frenesy), which quantifies the system’s jump rates with its environments. Originally derived for classical Markovian systems~\cite{Garrahan2017,Terlizzi2018, Yan2019, Hiura2021, Macieszczak2024}, KURs inherently rely on a particle-like transport picture valid in the weak system-bath coupling regime, where the concept of jumps (classical or quantum) is pertinent. This explains extensive recent studies of KURs in the quantum regime using a master equation approach valid for weak system-bath coupling~\cite{VanVu2022, VanVu2025, Prech2025, Liu2025, Liu2025b}. These studies addressed the interplay of KURs with TURs~\cite{Vo2022,Prech2023,Palmqvist2025}, and explored their validity in quantum coherent transport setups~\cite {Tesser2024, Bourgeois2024, Nishiyama2024, Prech2025}. Notably, within this regime, KURs were shown to yield tighter precision bounds compared to entropy-based uncertainty relations far from equilibrium. 

In the strong-coupling regime, quantum coherence and system-environment correlations blur individual transport events, challenging classical interpretations of activity. Establishing whether KURs hold in this regime has remained elusive due to the absence of an appropriate generalization of dynamical activity beyond weak-coupling. 
Here, equipped with our generalized definition of activity, we demonstrate clear violations of traditional KURs in strongly coupled quantum transport devices. To illustrate this explicitly, in Fig.~\ref{fig1} (lower panel) we show the SNR (solid black line) and the traditional KUR bound (dotted blue line) for a paradigmatic open quantum system: a single quantum dot (SQD) connected to two terminals under a voltage bias. 
% Despite the simplicity of this setup, 
We find clear violations of the standard KUR in the strong-coupling regime, defined by large values of the system–reservoir coupling strength $\Gamma$ with respect to the thermal energy scale $k_B T$.

%Objectives of This Paper
This state of current research motivates two fundamental questions:
(i) What is the physical meaning of dynamical activity beyond the weak-coupling regime?
(ii) Can KURs be derived in the quantum regime, valid at arbitrary coupling strengths and far from equilibrium?

To answer these questions, we analyze generic open quantum systems driven out of equilibrium by temperature and/or voltage biases. These systems consist of a central region coupled to multiple reservoirs via tunneling-type interactions $\hat{V}_\alpha$, see Fig.~\ref{fig1}. First, we introduce a generalized dynamical activity, defined from the zero-frequency spectral component of reservoir-system exchange rate fluctuations (Eq.~\eqref{eq:A_def}), valid at arbitrary coupling strengths. Explicit expressions are derived within two major theoretical frameworks for strong-coupling quantum transport: the Heisenberg equation of motion, valid at any time, and the Green’s function and Landauer–Büttiker formalisms in the stationary regime. We show that this generalized activity recovers the classical jump-based notion in the weak-coupling limit for a generic
system of $N$ coupled quantum dots described by a quadratic Hamiltonian.
Second, we establish a novel quantum kinetic uncertainty relation (QKUR) bounding the SNR as $I_\alpha^2/S_{\alpha \alpha}\leq \xi_{\mathrm{QKUR}}$. We illustrate these results with paradigmatic mesoscopic conductors: a quantum channel constrained by a quantum point contact (QPC) and single- and double-quantum dot systems. We demonstrate the validity of QKURs in all considered setups (see red dashed line in Fig.~\ref{fig1} for the SQD), and discuss the attainability of the bound.\\

\emph{Generalized dynamical activity.--} We consider an open quantum system where a fermionic multi-site system interacts with multiple fermionic reservoirs at equilibrium. The total Hamiltonian is decomposed as $\hat{H} = \hat{H}_S + \hat{H}_{\text{res}} + \hat{V}$
%\begin{equation}
%\label{eq:Htot}
    %\hat{H} = \hat{H}_S + \hat{H}_{\text{res}} + \hat{V}
%\end{equation}
where $\hat{H}_S$ describes the system with discrete energy levels, $\hat{H}_{\text{res}} = \sum_\alpha \hat{H}_\alpha$ represents the reservoirs, and $\hat{V} = \sum_\alpha \hat{V}_\alpha$ mediates particle exchanges between the system and each reservoir $\alpha$.

Within this framework, we introduce the generalized dynamical activity for reservoir $\alpha$ as  
\begin{equation}
\label{eq:A_def}
    \mathcal{A}_\alpha(t) = \frac{1}{2 \hbar^2} \int_{-t}^{t} \mathrm{d}\tau
    \left\langle\!\! 
\left\langle \left\{
            \hat{V}_\alpha(t), 
            \hat{V}_\alpha(t+\tau) 
        \right\}
    \right\rangle\!\! \right\rangle,
\end{equation}
with the covariance $\langle\!\langle XY \rangle\! \rangle \equiv \langle XY \rangle - \langle X \rangle \langle Y \rangle$, $\langle \cdot \rangle$ the quantum grand canonical ensemble average, and $\{\cdot, \cdot\}$ the anticommutator. 
% \gh{This definition highlights the physics captured by the generalized dynamical activity: it} characterizes the time-integrated and symmetrized fluctuations of tunneling processes between the system and the reservoirs. 
This quantity naturally captures coherent particle tunneling between system and reservoir through symmetrized fluctuations of the tunneling Hamiltonian, and reduces to the standard dynamical activity in the weak-coupling limit, as we demonstrate below.
% Importantly, Eq.~\eqref{eq:A_def} applies to arbitrary interaction Hamiltonians $\hat{V}_\alpha$.
Throughout this work, motivated by quantum transport applications, we focus on fermionic reservoirs coupled to the system via tunneling interactions of the form $\hat{V}_\alpha = \sum_{jk} \left( t_{jk\alpha}^* \hat{c}_{k\alpha}^\dagger \hat{d}_j + t_{jk\alpha} \hat{d}_j^\dagger \hat{c}_{k\alpha} \right)$. It governs tunneling between the $j$-th site of the system and the $k$-th mode of reservoir $\alpha$, with complex tunneling amplitudes $t_{jk\alpha}$. Each reservoir is described by its Hamiltonian 
$\hat{H}_\alpha = \sum_{k} \epsilon_{k\alpha} \hat{c}^\dagger_{k\alpha} \hat{c}_{k\alpha}$ 
with corresponding energy $\epsilon_{k\alpha}$ for mode $k$ in reservoir $\alpha$, 
with $\hat{c}_{k\alpha}$ and $\hat{c}^\dagger_{k\alpha}$ the fermionic annihilation and creation operators. The system's operators $\hat{d}_j^\dagger$ and $\hat{d}_j$ are the fermionic creation and annihilation operators for energy level $j$ of the system.

By employing the non-equilibrium Keldysh Green’s-function formalism, we derive an explicit expression
for the generalized dynamical activity in the steady state (denoted by the superscript $ss$). Following previous analytical developments by some of the authors in Ref.~\cite{Blasi2024} (see also Supp. Mat. Sec.~III~\cite{SuppMat}), we obtain:
\begin{align}
\label{eq_main:activityGreenmultiterminal}
\mathcal{A}^{ss}_\alpha = \frac{1}{\hbar}
\int \frac{d\epsilon}{2\pi} \Big\{&
\sum_{\beta\neq\alpha}
\Tr\!\left[ \mathbf{T}_{\alpha\beta}(\epsilon) \right]
\big(F_{\alpha\beta}(\epsilon) + F_{\beta\alpha}(\epsilon)\big)
\nonumber\\
&+
\Tr\Big[ 4\mathbf{T}_{\alpha\alpha}(\epsilon)
- \big(\sum_{\beta} \mathbf{T}_{\alpha\beta}(\epsilon)\big)^2 \Big]
F_{\alpha\alpha}(\epsilon)
\Big\}.
\end{align}
Here the matrices $\mathbf{T}_{\alpha\beta}(\epsilon)
=\boldsymbol{\Gamma}_\alpha(\epsilon)\mathbf{G}^r(\epsilon)
\boldsymbol{\Gamma}_\beta(\epsilon)\mathbf{G}^a(\epsilon)$
are defined in terms of the system's retarded and advanced Green's functions,
$\mathbf{G}^r(\epsilon)$ and
$\mathbf{G}^a(\epsilon)=\left[\mathbf{G}^r(\epsilon)\right]^\dagger$, with coupling
matrices
$[\boldsymbol{\Gamma}_\alpha(\epsilon)]_{ij}
=2\pi\sum_k t_{ik\alpha}t^*_{jk\alpha}\delta(\epsilon-\epsilon_{k\alpha})$
(see Refs.~\cite{Meir1992,Crepieux2024}).
The statistical factors
$F_{\alpha\beta}(\epsilon)=f_\alpha(\epsilon)\bigl[1-f_\beta(\epsilon)\bigr]$
are expressed in terms of the Fermi--Dirac distribution
$f_{\alpha}(\epsilon)=\{e^{(\epsilon-\mu_{\alpha})/k_B T_{\alpha}}+1\}^{-1}$,
with temperature $T_\alpha$ and chemical potential $\mu_\alpha$. In the following, we consider the wide-band limit with local couplings for comparison with the master-equation benchmark discussed below, such that
$\boldsymbol{\Gamma}_\alpha(\epsilon)=\Gamma_\alpha\,\mathbf{\Pi}_\alpha$,
where $\Gamma_\alpha$ is an energy-independent tunneling rate and $\mathbf{\Pi}_\alpha$ denotes the projector matrix to the system site coupled to reservoir $\alpha$. \\

%Below, we establish the connection between the generalized dynamical activity defined at all times in Eq.~\eqref{eq:A_def} and in the steady state in Eq.~\eqref{eq_main:activityGreenmultiterminal} and the standard jump-rate activity in the weak-coupling regime. \\

\textit{Recovery of the standard activity in the weak-coupling limit.--} 
A key consistency check for the generalized activity is that it reproduces the standard jump-rate activity in the weak-coupling regime within a master-equation (ME) approach. Within the ME framework, particle exchange with reservoir $\alpha$ is described by quantum jumps that add or remove a fermion on the locally coupled site. The corresponding activity, defined as the total rate of such events, is defined as~\cite{LandiPRXQuantum2024,Blasi2024}:
\begin{align}
A^{\rm ME}_\alpha (t)
&=\frac{1}{\hbar}\Tr\!\left[(\mathcal{L}^+_\alpha+\mathcal{L}^-_\alpha)\rho(t)\right].
\label{eq:AME_def}
\end{align}
This activity corresponds to the total rate of quantum jumps generated by the superoperators $\mathcal{L}^\pm_\alpha$, \( \mathcal{L}^+_\alpha\rho=\Gamma_\alpha f_\alpha \hat{d}_\alpha^\dagger\rho \hat{d}_\alpha, \) and \( \mathcal{L}^-_\alpha\rho=\Gamma_\alpha(1-f_\alpha)\hat{d}_\alpha\rho \hat{d}_\alpha^\dagger \),  where $\hat{d}_\alpha^\dagger$ ($\hat{d}_\alpha$) creates (annihilates) a fermion on the system site
coupled to reservoir $\alpha$, and $\Gamma_\alpha$ is the tunneling rate.
The occupation factor $f_\alpha$ corresponds to the Fermi function of reservoir $\alpha$,
evaluated at the transition energy of the dot level coupled to that reservoir.

For a single quantum dot (SQD), the correspondence between the two expressions Eq.~\eqref{eq:A_def} and Eq.~\eqref{eq:AME_def} can be formally established at all times analytically 
by exploiting an exact solution of the Heisenberg equations of motion for this system \cite{Blasi2024}. The proof is provided in Supp. Mat. Sec.~I~\cite{SuppMat}. 
For a generic system of $N$ quantum dots described by a quadratic Hamiltonian,
$\hat H_S=\sum_{ij} h_{ij} \hat{d}_i^\dagger \hat{d}_j$, the same correspondence can be
demonstrated in the steady state regime, considering the limit
$t \rightarrow \infty$ in Eq.~\eqref{eq:AME_def}. The proof exploits a
covariant formulation of Eq.~\eqref{eq:AME_def} that we can further express
within the formalism of Green's functions:
\begin{align}
A^{\rm ME}_\alpha
=\frac{1}{\hbar}\Big[\Gamma_\alpha f_\alpha+\Gamma_\alpha(1-2f_\alpha)\Tr(\mathbf{\Pi}_\alpha \mathbf{C}^{\rm ss})\Big],
\label{eq:AME_cov}
\end{align}
with the steady-state covariance matrix $\mathbf{C}^{\rm ss}$ and elements
$C^{\rm ss}_{ij}=\mathrm{Tr}[\hat{d}_j^\dagger \hat{d}_i\rho_{\rm ss}]$.
In this case, $\mathbf C^{\rm ss}$ obeys a closed linear Lyapunov
equation~\cite{Landi2022},
\begin{equation}
  \dot{\mathbf C}
  =
  \mathbf K\,\mathbf C
  +
  \mathbf C\,\mathbf K^\dagger
  +
  \mathbf P,
  \qquad
  \mathbf K \equiv -i\,\mathbf h_{\mathrm{eff}} ,
  \label{eq:Lyap-eom}
\end{equation}
where $\mathbf h_{\mathrm{eff}}=\mathbf h-\tfrac{i}{2}\sum_\beta \Gamma_\beta
\boldsymbol{\Pi_\beta}$ is the effective Hamiltonian in the wide-band limit,
and $\mathbf P=\sum_\beta \Gamma_\beta f_\beta\,\mathbf\Pi_\beta$.
Imposing $\dot{\mathbf C}=0$, the steady-state covariance matrix admits the formal solution~\cite{Meier2026}
\begin{equation}
  \mathbf C^{\rm ss}
  =
  \int_0^{\infty}\! \dd t\;
  e^{\mathbf K t}\,\mathbf P\,e^{\mathbf K^\dagger t}
  =
  \int\!\frac{\dd \epsilon}{2\pi}\;
  \mathbf G^r(\epsilon)\,\mathbf P\,\mathbf G^a(\epsilon),
  \label{eq:Css_GF}
\end{equation}
where, in the second equality, we used the identity
$\int_0^{\infty}\!\dd t\,e^{(\mathbf K+i\epsilon)t}
=
i[\,\epsilon\mathds 1-i\mathbf K\,]^{-1}
=
i\,\mathbf G^r(\epsilon)$ and $\int \frac{\dd \epsilon}{2\pi}e^{i \epsilon t}=\delta(t)$. 
Substituting Eq.~\eqref{eq:Css_GF} into Eq.~\eqref{eq:AME_cov}, and using Dyson’s identity
$\mathbf G^r(\epsilon)
\left(\sum_\beta \Gamma_\beta \mathbf\Pi_\beta\right)
\mathbf G^a(\epsilon)
=
i(\mathbf G^r(\epsilon)-\mathbf G^a(\epsilon))$
together with the spectral sum rule
$\int_{-\infty}^{\infty}\frac{d\epsilon}{2\pi}
\,i\big(\mathbf{G}^r_{\alpha\alpha}(\epsilon)-\mathbf{G}^a_{\alpha\alpha}(\epsilon)\big)=1$,
one finally obtains
\begin{align}
A^{\rm ME}_\alpha
=\frac{1}{\hbar}&\int\!\frac{d\epsilon}{2\pi}
\Big\{\sum_{\beta\neq\alpha}
\Tr\!\left[\mathbf{T}_{\alpha\beta}(\epsilon)\right]
\bigl[f_\alpha(1-f_\beta)+f_\beta(1-f_\alpha)\bigr]
\nonumber\\
&\quad
+2\Tr\!\left[\mathbf{T}_{\alpha\alpha}(\epsilon)\right]f_\alpha(1-f_\alpha)
\Big\}.
\label{eq:AME_final}
\end{align}
In the End Matter, we provide the final step of the demonstration, showing that Eq.~\eqref{eq:AME_final} corresponds to the weak-coupling limit of the generalized steady-state activity $\mathcal{A}^{ss}_\alpha$ in Eq.~\eqref{eq_main:activityGreenmultiterminal}.
%In the weak-coupling regime, the level broadening $\Gamma$ sets the smallest energy scale controlling transport, so that the Fermi functions vary smoothly on this scale and can be taken as approximately constant over the relevant energy window. Under this condition, one can directly compare the two expressions, including the quadratic combination of transmission matrices $\mathbf{T}_{\alpha\beta}(\epsilon)$ appearing in the Green’s-function formulation and absent in the master-equation approach.

Importantly, this result holds for a generic system of $N$ coupled quantum dots described by an
arbitrary quadratic Hamiltonian, independently of the internal structure of the system.
See the Supp. Mat.~\cite{SuppMat} for the explicit calculation for a double quantum dot (DQD).\\
%As an explicit illustration, in the Supplemental Material~\cite{SuppMat} we present the full calculation for a double quantum dot (DQD) system, where the equivalence between the two approaches is verified by taking the formal weak-coupling limit $\Gamma\to0$, corresponding to the Born approximation, while keeping $\Gamma t$ finite, which ensures the Markov approximation, as discussed in Ref.~\cite{Blasi2024}.\\

%%%%%%%%%%%%%%%%%%%%%%%%%%%%%%%%%%%%%%%%

\textit{Breakdown of standard KURs and quantum coherent contributions to $\mathcal{A}^{ss}_\alpha$.—}
As a paradigmatic example, we consider the SQD and show in Fig.~\ref{fig1} the KUR bound based on the generalized activity together with the SNR. Although the KUR holds at weak coupling ($\Gamma/k_B T \lesssim 2$),
it clearly breaks down at strong coupling.
To understand the origin of this breakdown and construct a quantum extension valid at arbitrary coupling strengths, we analyze the structure of the steady-state activity.
In the following we decompose $\mathcal{A}^{ss}_\alpha$ into contributions reflecting different physical processes, which form the basis of the QKUR derived below.

Equation~\eqref{eq_main:activityGreenmultiterminal} contains two contributions
proportional to the statistical factor
$F_{\alpha\beta}(\epsilon)=f_\alpha(\epsilon)[1-f_\beta(\epsilon)]$.
One is proportional to $F_{\alpha\alpha}(\epsilon)$ and corresponds to
auto-correlated events in reservoir $\alpha$, while the other involves terms
with $\beta\neq\alpha$, describing cross-correlated processes between
different reservoirs:
\begin{equation}
\label{eqmain:A_auto_cross}
\mathcal{A}^{ss}_\alpha=\mathcal{A}_{\alpha}^{auto}+\mathcal{A}_{\alpha}^{cross}\,.
\end{equation}
By exploiting the relation $F_{\alpha\beta}(\epsilon)+F_{\beta\alpha}(\epsilon)=(f_\alpha(\epsilon)-f_\beta(\epsilon))^2+F_{\alpha\alpha}(\epsilon)+F_{\beta\beta}(\epsilon)$, Eq.~\eqref{eqmain:A_auto_cross} can be recast into: 
\begin{equation}
\label{eqmain:A_th_shot}
\mathcal{A}^{ss}_\alpha=\mathcal{A}_{\alpha}^{th}+\mathcal{A}_{\alpha}^{sh}\,.
\end{equation}
Here $\mathcal{A}_\alpha^{th}$ denotes the thermal contribution. It vanishes in the zero‑temperature limit and remains finite at equilibrium. The other term, $\mathcal{A}_\alpha^{sh}$, is the nonequilibrium shot contribution. The decomposition in Eq.~\eqref{eqmain:A_th_shot} mirrors the one of the current noise in quantum transport (see \cite{Blanter2000} for a review).
% , which was key for demonstrating genuine quantum features such as the Fractional Quantum Hall regime~\cite{Saminadayar97, Heiblum98, Sabo_2017}. 
%All contributions in Eqs.~\eqref{eqmain:A_auto_cross} and~\eqref{eqmain:A_th_shot} will play a fundamental role in deriving a QKUR at strong coupling as discussed below.} 
%The decomposition of the steady-state generalized activity into distinct contributions, as shown in Eqs.~\eqref{eqmain:A_auto_cross} and~\eqref{eqmain:A_th_shot}, will play a crucial role in our discussion of the quantum KUR bound below. We emphasize that such a decomposition closely mirrors the structure familiar from the analysis of current noise $S_{\alpha\alpha}$ discussed in Ref.~\cite{Blanter2000}. 
%Equation~\eqref{eq_main:activityGreenmultiterminal} provides a useful definition of the generalized dynamical activity within a Hamiltonian's approach based on Green's functions. In the context of quantum coherent transport, it becomes relevant to express it within a Landauer-B\"uttiker approach.
%To provide a complete formalism for assessing the generalized dynamical activity in the steady-state regime, we derive its expression within the scattering-matrix (Landauer-Büttiker) framework. 
Using the Fisher-Lee relation \cite{Fisher1981,Lopez2004,Sanchez2025} which connects the retarded Green’s function to the scattering matrix, we show in Sec. IV of the Supp. Mat.~\cite{SuppMat} that the thermal and shot noise contributions defined in Eq.~\eqref{eqmain:A_th_shot} take the form:
\begin{align}
\label{eq:Fisher_Lee_thermal}
\mathcal{A}_{\alpha}^{th} =& \frac{1}{\hbar}\int \frac{d\epsilon}{2\pi}\Big\{ \left[1+\mathcal{R}_{\alpha\alpha}(\epsilon)\left(1-2\cos{(\phi_\alpha)}\right)\right] F_{\alpha\alpha}(\epsilon)\nonumber\\
&+\sum_{\beta\neq \alpha} \mathcal{T}_{\alpha\beta}(\epsilon)F_{\beta\beta}(\epsilon)\Big\}\,; \\
\label{eq:Fisher_Lee_shot}
\mathcal{A}_{\alpha}^{sh} =&  \frac{1}{\hbar} \sum_{\beta\neq \alpha}\int \frac{d\epsilon}{2\pi} \mathcal{T}_{\alpha\beta}(\epsilon) \left(f_\alpha(\epsilon)-f_\beta(\epsilon)\right)^2,
\end{align}
while the auto- and cross-parts read: 
\begin{align}
\label{eq:Fisher_Lee_auto}
\mathcal{A}_{\alpha}^{auto} &=  \frac{2}{\hbar}\int \frac{d\epsilon}{2\pi} \mathcal{R}_{\alpha\alpha}(\epsilon)\left(1-\cos{(\phi_\alpha)}\right) F_{\alpha\alpha}(\epsilon);\\
\label{eq:Fisher_Lee_cross}
\mathcal{A}_{\alpha}^{cross} &= \frac{1}{\hbar} \sum_{\beta\neq \alpha}\int \frac{d\epsilon}{2\pi} \mathcal{T}_{\alpha\beta}(\epsilon) \left(F_{\alpha\beta}(\epsilon)+F_{\beta\alpha}(\epsilon)\right).
\end{align}
These expressions explicitly depend on the scattering-matrix elements $s_{\alpha\beta}$ through the reflection and transmission probabilities of the quantum coherent mesoscopic conductor: $\mathcal{R}_{\alpha\alpha} = \abs{s_{\alpha\alpha}}^2$ (reflection into reservoir $\alpha$) and $\mathcal{T}_{\alpha\beta} = \abs{s_{\alpha\beta}}^2$ (transmission from reservoir $\beta$ to $\alpha$). The phase $\phi_\alpha$ appears in the complex reflection amplitude $r_{\alpha\alpha} = \sqrt{\mathcal{R}_{\alpha\alpha}} e^{i\phi_\alpha/2}$.

Interestingly, the thermal- and auto-contributions, Eqs.~\eqref{eq:Fisher_Lee_thermal} and~\eqref{eq:Fisher_Lee_auto}, remain finite even when the system is coupled to a single reservoir, as they explicitly depend on the local reflection probability $\mathcal{R}_{\alpha\alpha}$. In contrast, the shot- and cross-terms, Eqs.~\eqref{eq:Fisher_Lee_shot} and~\eqref{eq:Fisher_Lee_cross}, vanish in the absence of transmission, as they capture cross-correlated events between distinct reservoirs. This structural distinction is crucial for the form of the Quantum KUR derived below.\\
%%%%%%%%%%%%%%%%%%%%%%%%%%%%%%%%%%%%

\textit{Quantum KUR.--} 
%As recalled in the introduction and shown in Fig.~\ref{fig1} for the case of a SQD, KURs expressed in terms of activity are violated at strong-coupling. 
Our result for the generalized dynamical activity, Eq.~\eqref{eq_main:activityGreenmultiterminal},
together with its decomposition into cross and shot contributions,
Eqs.~\eqref{eq:Fisher_Lee_shot} and \eqref{eq:Fisher_Lee_cross}, allows us to derive a new bound for the
SNR:
\begin{equation}
\label{eq:QKUR}
    \frac{I_{\alpha}^2}{S_{\alpha\alpha}} \leq  \frac{(\mathcal{A}_{\alpha}^{cross})^2}{\mathcal{A}_{\alpha}^{cross} - \mathcal{A}_{\alpha}^{sh}}\equiv \xi_{{\rm QKUR}}\,.
\end{equation} 
We refer to this bound as the Quantum Kinetic Uncertainty Relation (QKUR), which we prove in the End Matter. 
Remarkably, the bound is entirely expressed in terms of the contributions to the generalized dynamical activity associated with cross-correlated processes between distinct reservoirs, $\mathcal{A}_\alpha^{\text{cross}}$ and $\mathcal{A}_\alpha^{\text{sh}}$. This reflects the necessity of having at least two reservoirs to obtain a nonvanishing SNR. The QKUR also emphasizes the crucial role of the shot contribution in the denominator: at low temperatures, $\mathcal{A}_\alpha^{\text{sh}}$ increases, causing $\xi_{\mathrm{QKUR}}$ to grow and ensuring the validity of Eq.~\eqref{eq:QKUR} even at strong-coupling.

Another limit of interest corresponds to thermal equilibrium (zero voltage bias and equal temperatures for the reservoirs), where $\mathcal{A}_\alpha^{\text{sh}}$ vanishes and Eq.~\eqref{eq:QKUR} reduces to the bound derived in Refs.~\cite{Splettstoesser2024, Palmqvist2025}. 
The relation reported in Ref.~\cite{Splettstoesser2024} is of a fundamentally different nature from our result. Their bound applies only to the classical signal-to-noise ratio $I^2/S^{cl}$ and is derived without a consistent definition of dynamical activity for coherent mesoscopic transport. 
The quantity introduced there, while reminiscent of an activity, corresponds in fact only to the cross part of the generalized activity; such a bound is necessarily restrictive, since including the shot contribution in the denominator of $\xi_{\mathrm{QKUR}}$ is essential to ensure validity far from equilibrium.
In this respect, the result of Ref.~\cite{Splettstoesser2024} is confined to the near-equilibrium, classical limit, while our QKUR provides the general bound for the full signal-to-noise ratio $I^2/S$, with earlier results recovered only as particular limiting cases.\\

%We now investigate and illustrate the validity of this new uncertainty relation QKUR for paradigmatic quantum-coherent mesoscopic conductors.\\

%%%%%%%%%%%%%%%%%%%%%%%%%%%%%%%%%%%%%

\begin{figure}[t]
{{\includegraphics[width=.5\textwidth]{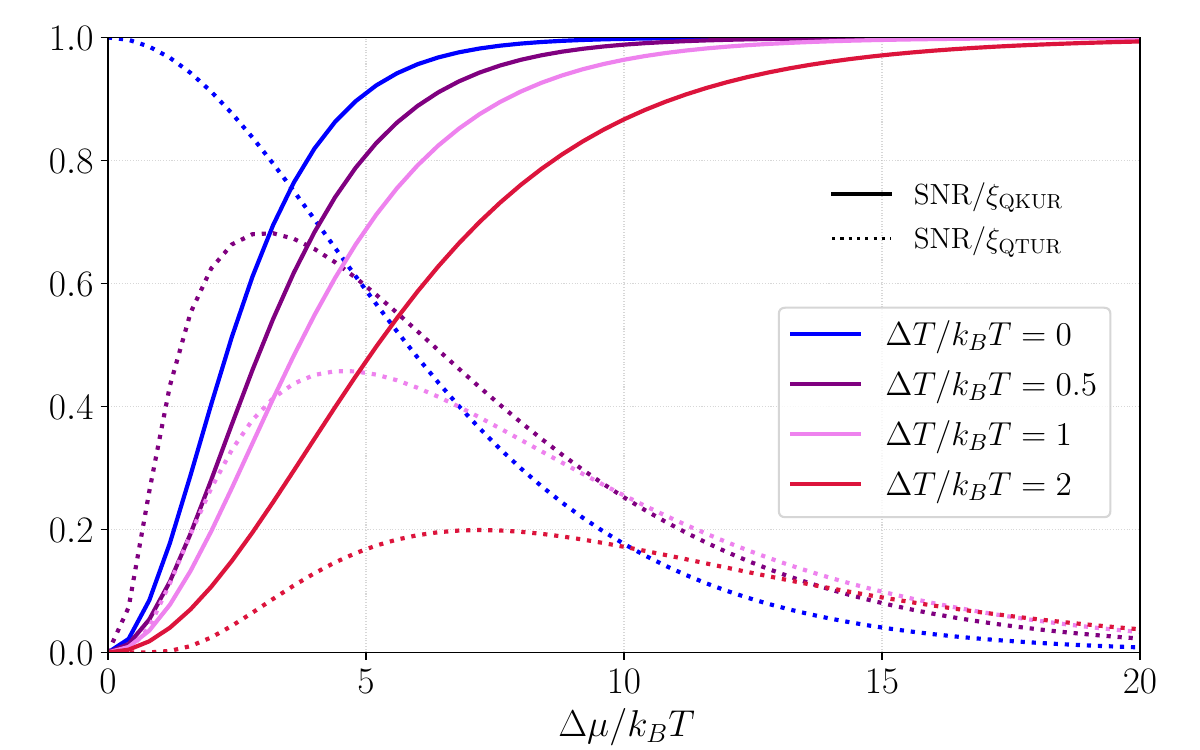}}}
\caption{Uncertainty relations for a perfectly transmitting QPC ($\tau = 1$). The ratios $\text{SNR}/\xi_{\mathrm{QKUR}}$ (solid) and $\text{SNR}/\xi_{\mathrm{QTUR}}$ (dashed)~\cite{Brandner2025} are shown as functions of voltage bias $\Delta\mu/k_B T$ for different thermal biases $\Delta T/k_B T$. Both QKUR and QTUR remain valid at arbitrary coupling strength. Notably, QKUR provides a tight bound far from equilibrium, with $\text{SNR}/\xi_{\mathrm{QKUR}} \rightarrow 1$ at large $\Delta\mu/k_B T$ for all $\Delta T$.}
\label{fig:Bound_QPC}
\end{figure}

\noindent\textit{QKUR for a Quantum Point Contact.--}  
We consider a Quantum Point Contact (QPC) connecting two reservoirs, modeled as a single quantum channel in the ballistic regime with an energy-independent transmission probability $\mathcal{T}_{\alpha \beta}(\epsilon) \equiv \tau$, and subject to both voltage and temperature biases.
% Analytical expressions for the generalized dynamical activity, the SNR, and the QKUR bound $\xi{\mathrm{QKUR}}$ follow directly from the general formulas.
Figure~\ref{fig:Bound_QPC} shows the results for a perfectly transmitting channel ($\tau = 1$), a case of particular interest for testing QKUR in the strong-coupling regime, where no semi-classical jump-based description applies. We plot the ratio ${\rm SNR}/\xi_{\mathrm{QKUR}}$ (solid lines) as a function of the rescaled voltage bias $\Delta \mu/k_B T$, for different values of the rescaled temperature bias $\Delta T/k_B T$.
The ratio ${\rm SNR}/\xi_{\mathrm{QKUR}}$ remains strictly below 1, confirming that $\xi_{\mathrm{QKUR}}$ provides a valid bound across all coupling regimes. Remarkably, all solid curves saturate to unity for $\Delta \mu \gg k_B T$, demonstrating that the QKUR bound becomes tight precisely in the far from equilibrium limit. 
For the sake of completeness, and to connect with most recent results in the field, we also plot the corresponding ratio ${\rm SNR}/\xi_{\text{QTUR}}$ (dashed curves), where $\xi_{\text{QTUR}} = I_\alpha \sinh{(\sigma/2k_B I_\alpha)}$ represents the bound for a Quantum TUR (QTUR) introduced in Eq.~(2) of Ref.~\cite{Brandner2025}. It depends on the entropy production rate $\sigma$ and the particle current $I_\alpha$. Interestingly, QTUR is tight near equilibrium ($\Delta \mu, \Delta T \approx 0$), where the ratio approaches unity, but becomes increasingly loose as the system moves away from equilibrium (e.g., for finite $\Delta \mu$ and $\Delta T$).
This comparison highlights the complementary nature of the two bounds: QKUR is optimal far from equilibrium, while QTUR performs best near equilibrium. %For completeness, a detailed analysis of the QKUR bound for a double quantum dot (DQD) system is provided in the End Matter.\\

%%%%%%%%%%%%%%%%%%%%%%%%%%%%%%%%%%%%%%%%%%%

\textit{Conclusions.--} 
Starting from a microscopic description, we introduced a generalized dynamical activity defined in terms of exchange rate fluctuations associated with the interaction Hamiltonian between a quantum system and its external reservoirs. In the steady state, we derived explicit expressions within both the Green's function and Landauer–Büttiker formalisms, suited for quantum coherent transport. We showed that this generalized activity reduces to the total jump rate in the weak-coupling limit, recovering the standard definition used in the master equation framework.

Building on this concept, we proposed a novel uncertainty bound for the signal-to-noise ratio, valid at arbitrary coupling strength. The corresponding uncertainty relation, denoted QKUR, was validated and its tightness analyzed in paradigmatic open quantum systems: a quantum channel pinched by a QPC, and single- and double-quantum dot setups, especially in far from equilibrium conditions under temperature and voltage biases.

Our work thus establishes the first general activity-based framework for uncertainty relations valid at arbitrary coupling. This framework reveals why the standard KUR breaks down and provides its natural quantum extension, the QKUR, in close analogy with the TUR–QTUR paradigm. 

While our analysis focuses on quadratic systems, this regime already captures the role of quantum coherence in transport and constitutes the standard setting for state-of-the-art approaches to uncertainty relations in mesoscopic conductors. Importantly, we show that the breakdown of the standard KUR already occurs within this analytically controlled framework, indicating that its failure is not tied to many-body interactions but rather to the limitations of the weak-coupling, jump-based description of dynamical activity. Extending the present approach to regimes involving strong many-body correlations represents an interesting direction for future work.
Another promising direction opened by this work is the extension of the
general framework presented here to QKURs for other transport quantities,
such as energy and heat currents. % In this work we focused on particle currents; however, the formalism is sufficiently general to be extended to other transport quantities, such as energy or heat currents, which we leave as a promising direction for future investigations.\\

%%%%%%%%%%%%%%%%%%%%%%%%%%%%%%%%%%%%%%%%%%%%%%%%%%%%%%%%%%%%

%TC:ignore
\textit{Acknowledgments.--} We acknowledge fruitful discussions with Micha{\l} Horodecki and Giovanni Maria Brandi for useful comments. G.B. and G.H. acknowledge support from the National Center of Competence and Research SwissMAP and G.H. acknowledges support from the FPFS - De Meuron program for academic promotion. R.R.R. is financially supported by the Conselleria d’Educació i Universitats del Govern de les Illes Balears and Fons Social Europeu+ through the contract POSTDOC2024\_17 and the Government of Spain (Severo Ochoa CEX2019-000910-S and FUNQIP), Fundació Cellex, Fundació Mir-Puig, Generalitat de Catalunya (CERCA program). M.M. acknowledges the support from CSIC/IUCRAN2022 under Grant No. UCRAN20029. G.B. and R.L. acknowledges support by the Spanish State Research Agency (MCIN/AEI/10.13039/501100011033), FEDER (UE) and ERDF (EU) under Grant No. PID2023-151975NB-I00.

%%%%%%%%%%%%%%%%%%%%%%%%%%%%%%%%%%%%%%%%%%%%%%%%%%%%%%%%%%%%

\bibliography{References}
\bibliographystyle{bibstyle}

%%%%%%%%%%%%%%%%%%%%%%%%%%%%%%%%%%%%%%%%%%%%%%%%%%%%%%%%%%%%

\clearpage
\newpage

\setcounter{equation}{0}
\renewcommand{\theequation}{EM\arabic{equation}}

%TC:ignore
\section*{End Matter}

\subsection{Proof of equivalence in the weak-coupling regime}
In this section, we show that the steady-state generalized dynamical activity reduces, in the weak-coupling regime, to the jump-rate activity obtained within a master-equation (ME) approach.
For quadratic Hamiltonians, both quantities can be expressed in terms of Green’s functions using the covariance matrix formalism. For clarity, we recall their steady-state expressions. The generalized activity reads
\begin{align}
\mathcal{A}^{ss}_\alpha =&
\frac{1}{\hbar}
\int \frac{d\epsilon}{2\pi}
\Big\{
\sum_{\beta\neq\alpha}
\Tr\!\left[ \mathbf{T}_{\alpha\beta}(\epsilon) \right]
\big(F_{\alpha\beta}(\epsilon) + F_{\beta\alpha}(\epsilon)\big)\nonumber\\
&+
\Tr\!\Big[
4\mathbf{T}_{\alpha\alpha}(\epsilon)
-
\big(\sum_{\beta} \mathbf{T}_{\alpha\beta}(\epsilon)\big)^2
\Big]
F_{\alpha\alpha}(\epsilon)
\Big\}.
\label{eq:EM_Ass}
\end{align}
The corresponding master-equation activity is
\begin{align}
A^{\rm ME}_\alpha =&
\frac{1}{\hbar}
\int\frac{d\epsilon}{2\pi}
\Big\{
\sum_{\beta\neq\alpha}
\Tr\!\left[\mathbf{T}_{\alpha\beta}(\epsilon)\right]
(F_{\alpha \beta} + F_{\beta \alpha})\nonumber\\
&+
2\Tr\!\left[\mathbf{T}_{\alpha\alpha}(\epsilon)\right]
F_{\alpha \alpha}
\Big\}.
\label{eq:EM_AME}
\end{align}
% Comparing Eqs.~\eqref{eq:EM_Ass} and \eqref{eq:EM_AME}, two differences must be addressed:
% (i) the energy dependence of the statistical factors, and
% (ii) the quadratic structure of the diagonal transmission term.
Comparing Eqs.~\eqref{eq:EM_Ass} and \eqref{eq:EM_AME}, two differences must be addressed.
First, in Eq.~\eqref{eq:EM_Ass} the statistical factors $F_{\alpha\beta}(\epsilon)=f_\alpha(\epsilon)\big(1-f_\beta(\epsilon)\big)$, depend explicitly on energy, whereas in Eq.~\eqref{eq:EM_AME} they appear as energy-independent quantities.
Second, the diagonal contribution in Eq.~\eqref{eq:EM_Ass} contains the quadratic structure $(\sum_\beta \mathbf{T}_{\alpha\beta})^2$, while in Eq.~\eqref{eq:EM_AME} it enters only linearly through $2\mathbf{T}_{\alpha\alpha}$.
\\
\paragraph*{Energy dependence of the statistical factors.}
In Eq.~\eqref{eq:EM_Ass}, the statistical factors depend explicitly on energy through the Fermi–Dirac distributions. In contrast, in Eq.~\eqref{eq:EM_AME}, they are energy independent. 
\\
In the weak-coupling regime, the level broadening $\Gamma$ is the smallest energy scale. In the wide-band limit, the Fermi functions vary smoothly on this scale and can therefore be treated as constant over the relevant integration window. As discussed in the main text, they are effectively evaluated at the transition energies of the dot levels coupled to the reservoirs. Under this weak-coupling approximation, the statistical factors in Eq.~\eqref{eq:EM_Ass} become energy independent and coincide with those in Eq.~\eqref{eq:EM_AME}. 
With this simplification, the first line of Eq.~\eqref{eq:EM_Ass} matches exactly the first line of Eq.~\eqref{eq:EM_AME}.
\\
\paragraph*{Quadratic structure of the transmission matrices.}
The remaining difference concerns the diagonal contribution proportional to $F_{\alpha\alpha}$. In Eq.~\eqref{eq:EM_Ass}, this term contains the quadratic combination $(\sum_\beta \mathbf{T}_{\alpha\beta}(\epsilon))^2$, whereas in Eq.~\eqref{eq:EM_AME} it appears only linearly as $2\mathbf{T}_{\alpha\alpha}(\epsilon)$. Equating the terms proportional to $F_{\alpha\alpha}$ in Eqs.~\eqref{eq:EM_Ass} and \eqref{eq:EM_AME}, the comparison reduces to proving the identity
\begin{equation}
\int\frac{d\epsilon}{2\pi}
\Tr\!\Big[
\Big(\sum_{\beta}\mathbf T_{\alpha\beta}(\epsilon)\Big)^2
\Big]
=
2
\int\frac{d\epsilon}{2\pi}
\Tr\!\big[\mathbf T_{\alpha\alpha}(\epsilon)\big].
\label{eq:EM_identity}
\end{equation}
Using $\mathbf T_{\alpha\beta}
=\boldsymbol{\Gamma}_\alpha
\mathbf G^r
\boldsymbol{\Gamma}_\beta
\mathbf G^a$
together with Dyson’s identity
$\mathbf G^r
\Big(\sum_\beta \boldsymbol{\Gamma}_\beta\Big)
\mathbf G^a
=
i(\mathbf G^r-\mathbf G^a)$, one obtains
\begin{equation}
\sum_\beta \mathbf T_{\alpha\beta}
=
i\boldsymbol{\Gamma}_\alpha(\mathbf G^r-\mathbf G^a).
\label{eq:EM_sumT}
\end{equation}
For local coupling to reservoir $\alpha$, with
$\boldsymbol{\Gamma}_\alpha=\Gamma_\alpha \mathbf\Pi_\alpha$
and  $\mathbf\Pi_\alpha$ the projector onto the coupled site, one finds
\begin{equation}
\Tr\!\Big[
\Big(\sum_\beta \mathbf T_{\alpha\beta}\Big)^2
\Big]
=
\Gamma_\alpha^2
\big[2\,\Im\,\mathbf G^r_{\alpha\alpha}(\epsilon)\big]^2.
\label{eq:EM_trace_square}
\end{equation}
Similarly, the diagonal transmission reads
\begin{equation}
\Tr\!\big[\mathbf T_{\alpha\alpha}(\epsilon)\big]
=
\Gamma_\alpha^2
\big|\mathbf G^r_{\alpha\alpha}(\epsilon)\big|^2.
\label{eq:EM_trace_linear}
\end{equation}
The problem thus reduces to relating the integrals of
$|\mathbf G^r_{\alpha\alpha}(\epsilon)|^2$
and
$(\Im\,\mathbf G^r_{\alpha\alpha}(\epsilon))^2$.
Since $\mathbf G^r_{\alpha\alpha}(\epsilon)$ is a retarded Green’s function, it is analytic in the upper half-plane, its real and imaginary parts obey Kramers--Kronig relations, yielding
\begin{equation}
\int_{-\infty}^{\infty} d\epsilon\;
\big|\mathbf G^r_{\alpha\alpha}(\epsilon)\big|^2
=
2
\int_{-\infty}^{\infty} d\epsilon\;
\big(\Im\,\mathbf G^r_{\alpha\alpha}(\epsilon)\big)^2,
\label{eq:EM_Hilbert}
\end{equation}
which establishes Eq.~\eqref{eq:EM_identity}.
\\
This establishes the equivalence between Eqs.~\eqref{eq:EM_Ass} and \eqref{eq:EM_AME} in the weak-coupling steady-state limit, showing that the generalized dynamical activity consistently recovers the standard jump-rate definition.

%%%%%%%%%%%%%%%%%%%%%%%%%%%%%%%%%%%%%%%%%%%%%%%%%%%%%%%%%%%%%%%%%%%%%%%%%%%%%%%%%%%%%%%

\subsection{Proof for the QKUR bound}
In this section, we derive bounds on the signal-to-noise ratio in terms of the stationary generalized activity for a generic multi-terminal system. 
We begin by establishing an upper bound for the current: 
\begin{align}
\label{eq:Ia}
\lvert I_{\alpha}\rvert
&=
\left\lvert
\frac{1}{\hbar}
\sum_{\beta\neq\alpha}
\int\frac{d\epsilon}{2\pi}\,
\mathcal T_{\alpha\beta}(\epsilon)
\bigl[f_{\alpha}(\epsilon)-f_{\beta}(\epsilon)\bigr]
\right\rvert\nonumber
\\
&\leq
\frac{1}{\hbar}
\sum_{\beta\neq\alpha}
\int\frac{d\epsilon}{2\pi}\,
\mathcal T_{\alpha\beta}(\epsilon)
\left\lvert
f_{\alpha}(\epsilon)-f_{\beta}(\epsilon)
\right\rvert\nonumber
\\
&\leq
\frac{1}{\hbar}
\sum_{\beta\neq\alpha}
\int\frac{d\epsilon}{2\pi}\,
\mathcal T_{\alpha\beta}(\epsilon)
\bigl[
F_{\alpha\beta}(\epsilon)+F_{\beta\alpha}(\epsilon)
\bigr]
=
A_{\alpha}^{\mathit{cross}}.
\end{align}
where in the last inequality we used $F_{\alpha\beta}+F_{\beta\alpha}\geq \abs{f_\alpha-f_\beta}\geq \left(f_\alpha-f_\beta\right)$. Similarly, for the noise we have 
\begin{align}
\label{eq:noise_Saa}
    S_{\alpha\alpha}&=\frac{1}{\hbar}\sum_{\beta\neq \alpha}\int \frac{d\epsilon}{2\pi} \mathcal{T}_{\alpha\beta}(\epsilon)\left(F_{\alpha\beta}(\epsilon)+F_{\beta\alpha}(\epsilon)\right)\nonumber\\
    &-\frac{1}{\hbar}\int \frac{d\epsilon}{2\pi}\left(\sum_{\beta\neq \alpha}\mathcal{T}_{\alpha\beta}(\epsilon)\left(f_\alpha(\epsilon)-f_\beta(\epsilon)\right)\right)^2\nonumber\\
    &\equiv S_{\alpha\alpha}^{cl}-S_{ \alpha\alpha}^{qu}=\mathcal{A}_{\alpha}^{cross}-S_{ \alpha\alpha}^{qu},
\end{align}
where in the last equality we introduced the decomposition of the noise into classical and quantum contributions, $S_{\alpha\alpha}=S^{cl}_{\alpha\alpha}-S^{qu}_{\alpha\alpha}$, corresponding respectively to the terms linear and quadratic in the transmission probabilities, and used that $\mathcal{A}_{\alpha}^{cross}=S_{\alpha\alpha}^{cl}$~\cite{SuppMat, Splettstoesser2024}.

To proceed, we now derive an upper bound for the quantum part of the noise, $S_{\alpha\alpha}^{qu}$. Applying the Cauchy–Schwarz inequality and using $\sum_{\beta\neq \alpha} \mathcal{T}_{\alpha\beta}(\epsilon) = 1 - \mathcal{R}_{\alpha\alpha} \leq 1$, we obtain:

\begin{align}  S_{\alpha\alpha}^{qu}&=\frac{1}{\hbar}\int \frac{d\epsilon}{2\pi}\left(\sum_{\beta\neq \alpha}\mathcal{T}_{\alpha\beta}(\epsilon)\left(f_\alpha(\epsilon)-f_\beta(\epsilon)\right)\right)^2\nonumber\\
&\leq \frac{1}{\hbar}\int \frac{d\epsilon}{2\pi}\left[\sum_{\beta\neq \alpha}\mathcal{T}_{\alpha\beta}(\epsilon)\left(f_\alpha(\epsilon)-f_\beta(\epsilon)\right)^2\right]\left[\sum_{\beta\neq \alpha}\mathcal{T}_{\alpha\beta}(\epsilon)\right]\nonumber\\
    &\leq \frac{1}{\hbar}\int \frac{d\epsilon}{2\pi}\sum_{\beta\neq \alpha}\mathcal{T}_{\alpha\beta}(\epsilon)\left(f_\alpha(\epsilon)-f_\beta(\epsilon)\right)^2\nonumber\\
    &=\mathcal{A}_{ \alpha}^{sh}\leq\mathcal{A}_{\alpha}^{cross}.
\end{align}
Using Eqs.~\eqref{eq:Ia} and \eqref{eq:noise_Saa}, together with the inequality
$S^{qu}_{\alpha\alpha} \le A^{sh}_\alpha$ established above,
we obtain an upper bound for the signal-to-noise ratio, given by
Eq.~\eqref{eq:QKUR} of the main text:
\begin{align}
    \frac{I_{\alpha}^2}{S_{\alpha\alpha}}\leq\frac{(\mathcal{A}_{\alpha}^{cross})^2}{\mathcal{A}_{\alpha}^{cross}-S_{\alpha\alpha}^{qu}}\leq \frac{(\mathcal{A}_{\alpha}^{cross})^2}{\mathcal{A}_{\alpha}^{cross}-\mathcal{A}_{\alpha}^{sh}}\equiv \xi_{{\rm QKUR}}.
\end{align}

\begin{figure}[t]
\includegraphics[width=0.5\textwidth]{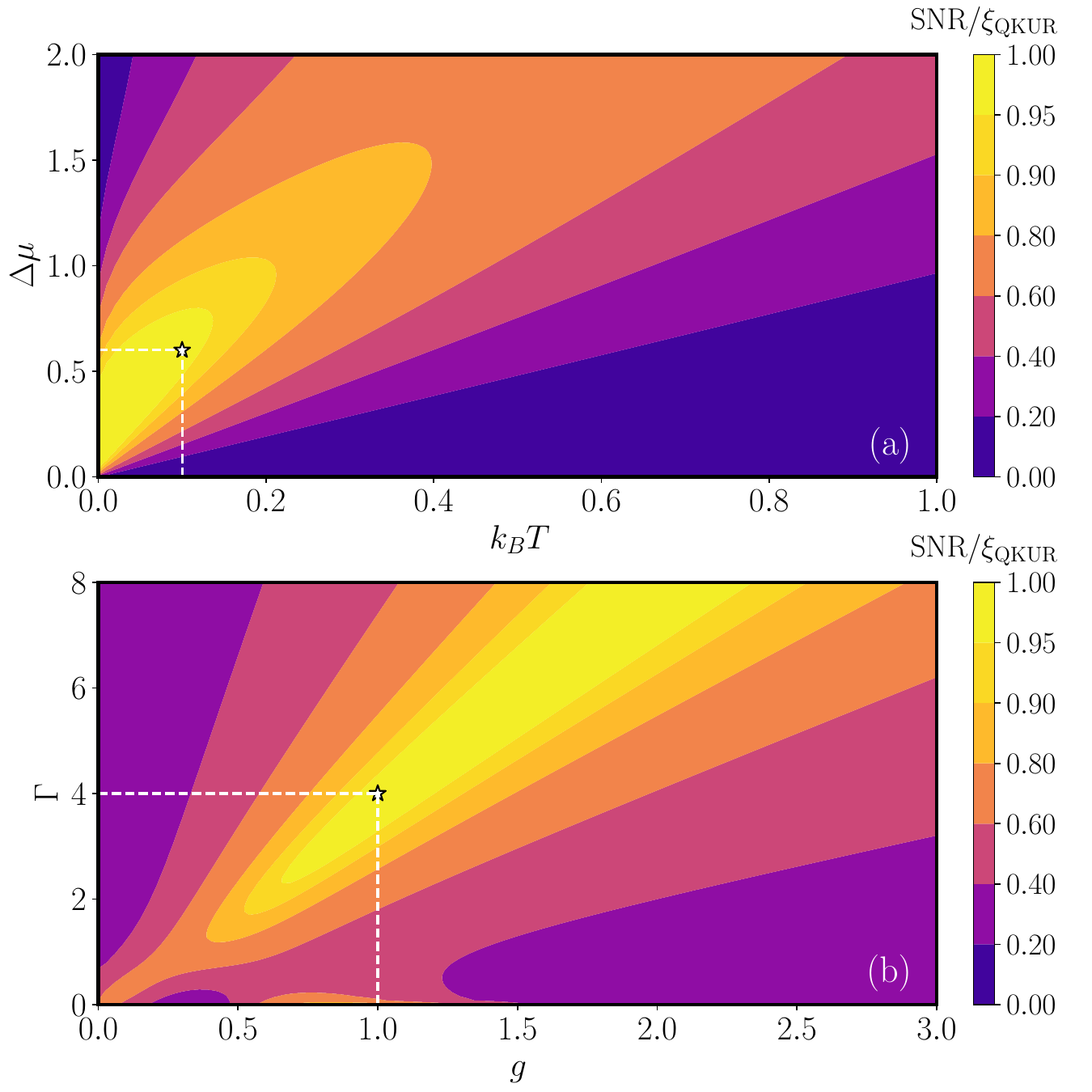}
\caption{Contour plots of the ratio ${\rm SNR}/\xi_\mathrm{QKUR}$ for the DQD as a function of $k_B T$ and $\Delta\mu$ (panel (a)), and as a function of the coupling strength $\Gamma_L = \Gamma_R=\Gamma/2$ and interdot tunneling $g$ (panel (b)). White stars mark the parameter values used in the complementary panel. An asymmetric voltage bias is applied such that $\Delta\mu = \mu_L - \mu_R$ with $\mu_R = 0$. No temperature bias is applied, and $\epsilon_1 = \epsilon_2 = 0$. }
\label{fig:QKUR_SQD_DQD}
\end{figure}

\subsection{QKUR for a Double Quantum Dot}

% As introduced earlier and illustrated in Fig.~\ref{fig1}, the QKUR has already been discussed for a single quantum dot (SQD), highlighting its validity in the strong-coupling regime where the standard KUR fails. 
% We now extend this analysis to another paradigmatic system: the double quantum dot (DQD). 
We now extend the analysis of the QKUR to another paradigmatic system: a double quantum dot (DQD).
The setup consists of two quantum dots connected to left ($L$) and right ($R$) reservoirs with tunneling rates $\Gamma_L$ and $\Gamma_R$. 
The system Hamiltonian is given by $\hat{H}_S = \sum_{n=1,2} \epsilon_n \hat{d}_n^{\dagger}\hat{d}_n + g \sum_{n\neq m} \hat{d}_n^{\dagger}\hat{d}_m$, where $\epsilon_1$ and $\epsilon_2$ denote the dot energies and $g$ the inter-dot tunneling amplitude.
In the following we consider the symmetric case $\epsilon_1=\epsilon_2=\epsilon_d$, corresponding to two energy-degenerate dots coupled in series.
% As introduced earlier and illustrated in Fig.~\ref{fig1}, the QKUR has already been discussed for a single quantum dot (SQD), highlighting its validity in the strong-coupling regime, where the standard KUR fails. We now extend this analysis to a second paradigmatic system: the double quantum dot (DQD). Both setups involve quantized energy-level systems in a two-terminal configuration, with left ($L$) and right ($R$) reservoirs coupled via tunneling rates $\Gamma_L$ and $\Gamma_R$. The DQD consists of two energy-degenerate dots at $\epsilon_1 = \epsilon_2 = \epsilon_d$, coupled in series through an inter-dot tunneling term of strength $g$. 
Each dot is locally connected to its respective reservoir, and the corresponding transmission probability is given by $\mathcal{T}_{RL}(\epsilon)=\mathcal{T}_{LR}(\epsilon)=
   g^2 \Gamma_L \Gamma_R/\left[\left(g^2+\frac{\Gamma_L\Gamma_R}{4}-(\epsilon-\epsilon_d)^2\right)^2
   +\frac{(\Gamma_L+\Gamma_R)^2}{4}(\epsilon-\epsilon_d)^2\right]$. 

Figure~\ref{fig:QKUR_SQD_DQD} shows the behavior of the ratio ${\rm SNR}/\xi_\mathrm{QKUR}$ as a function of the key parameters of the setup. In the top panel, we observe that the QKUR bound becomes tighter in the regime $\Delta \mu \gtrsim k_B T$, and in particular at low temperatures ($k_B T \lesssim \Gamma$), where a range of voltage biases leads to ${\rm SNR}$ values within 95\% of the bound (yellow regions). As in previous setups, the bound is saturated far from equilibrium in the strong-coupling regime.

In the bottom panel, we analyze ${\rm SNR}/\xi_\mathrm{QKUR}$ as a function of $\Gamma$ and the interdot coupling $g$, and find that the value of $\Gamma$ for which the bound is closest to the SNR increases approximately linearly with $g$.

%%%%%%%%%%%%%%%%%%%%%%%%%%%%%%%%%%%%%%%%%%%%%%

\clearpage
\newpage

% reset contatori
\setcounter{section}{0}
\setcounter{subsection}{0}

% riattiva numerazione
\setcounter{secnumdepth}{2}

% formato numerazione
\renewcommand{\thesection}{\Roman{section}}
\renewcommand{\thesubsection}{\thesection\ \Alph{subsection}}

\setcounter{equation}{0}
\renewcommand{\theequation}{SM\arabic{equation}}

\onecolumngrid

\begin{center}
{\large\bfseries SUPPLEMENTAL MATERIAL}
\end{center}

\vspace{0.5cm}

\section{Activity for Single Quantum Dot}
\label{app:SQD}
To validate the definition of dynamical activity introduced in Eq.~(2) of the main text, we benchmark it against known results in the weak-coupling regime for a single-level quantum dot (SQD) connected to two fermionic leads.
First, in Sec.~\ref{app_strong_coupling_activity_SQD}, we derive the expression for the activity for an SQD system in the strong-coupling regime using exact solutions of the Heisenberg equations.
Then, in Sec.~\ref{app_weak_coupling_protocol_SQD}, we show how to recover the analytical result obtained from the master equation approach in the weak-coupling regime, following the protocol introduced in Ref.~\cite{Blasi2024}.

\subsection{Strong-Coupling Activity for a SQD}
\label{app_strong_coupling_activity_SQD}

To compute the generalized activity in the strong-coupling regime, we begin with its definition in Eq.(2) of the main text, which can be rewritten as:
\begin{equation} \label{app_eq:Activity} \mathcal{A}_\alpha(t) = \frac{1}{\hbar^2} \int_{-t}^{t} d\tau \frac{\langle\langle \hat{V}_\alpha (t) \hat{V}_\alpha (t+\tau) \rangle\rangle + \langle\langle \hat{V}_\alpha (t + \tau) \hat{V}_\alpha (t) \rangle\rangle}{2} = \frac{1}{\hbar^2} \Re \int_{-t}^{t} d\tau~ \langle\langle \hat{V}_\alpha (t) \hat{V}_\alpha (t+\tau) \rangle\rangle, 
\end{equation}
where we used the identity $\langle\langle \hat{V}_\alpha (t + \tau) \hat{V}_\alpha (t) \rangle\rangle = \langle\langle \hat{V}_\alpha (t) \hat{V}_\alpha (t+\tau) \rangle\rangle^*$.
In the case of a SQD, the system Hamiltonian is $\hat{H}_S = \epsilon_d \hat{d}^{\dagger} \hat{d}$ (with $\epsilon_d$ the energy of the dot), and the tunneling interaction with lead $\alpha = L, R$ is described by: \begin{equation} \hat{V}_\alpha = \sum_{k} \left( t_{k\alpha}^{*} \hat{c}_{k\alpha}^{\dagger} \hat{d} + t_{k\alpha} \hat{d}^{\dagger} \hat{c}_{k\alpha} \right). \end{equation}
Using the above expression and applying Wick's theorem, we can write the argument of the real part in the integral of Eq. \eqref{app_eq:Activity} as
\begin{align}\label{eq:activity-after-Wick}
    \langle\langle \hat{V}_\alpha (t) \hat{V}_\alpha (t+\tau) \rangle\rangle &= \sum_{kk'} t^*_{k\alpha} t^*_{k'\alpha} \langle c_{k\alpha}^\dag(t) d(t+\tau) \rangle \langle d(t) c_{k'\alpha}^\dag(t+\tau)\rangle+   \sum_{kk'}t^*_{k\alpha} t_{k'\alpha} \langle c_{k\alpha}^\dag(t) c_{k'\alpha}(t+\tau) \rangle \langle d(t) d^\dag(t+\tau) \rangle \nonumber\\
    &+ \sum_{kk'} t_{k\alpha} t^*_{k'\alpha} \langle d^\dag(t) d(t+\tau)\rangle \langle c_{k\alpha}(t)c_{k'\alpha}^\dag(t+\tau)  \rangle +  \sum_{kk'} t_{k\alpha} t_{k'\alpha} \langle d^\dag(t) c_{k\alpha}(t+\tau) \rangle \langle c_{k\alpha}(t) d^\dag(t+\tau) \rangle.
\end{align}
Following Ref.~\cite{Blasi2024}, and using the Heisenberg equation of motion formalism and wide band limit (WBL) approximation, after a long calculation, the above expression takes the following form

\begin{align}
\label{app_eq:fluctuation_Lambda}
    \langle\langle \hat{V}_\alpha (t) \hat{V}_\alpha (t+\tau) \rangle\rangle &= 
    \sum_\beta \Gamma_\alpha \Gamma_\beta\left[ \Lambda_\alpha^{(0)}(t,t+\tau) \overline{\Lambda}_\beta^{(1)}(t+\tau,t) - \delta_{\alpha\beta} \Lambda^{(2)}_\alpha(t,t+\tau) \overline{\Lambda}_\alpha^{(2)}(t+\tau,t)\right]+\{\Lambda \leftrightarrow\overline{\Lambda}\}
\end{align}
where the adimensional $\Lambda$-functions are defined as

\begin{equation}
\label{Lambda_HE}
\begin{aligned}
&\Lambda^{(0)}_{\gamma}(t,t^{\prime}) = \frac{2}{\Gamma}\int \!\frac{d\epsilon}{2\pi} e^{i\left(\epsilon-\epsilon_d\right)\left(t-t^\prime\right)/\hbar} f_\gamma\left(\epsilon\right),\\
&\Lambda^{(1)}_{\gamma}(t,t^{\prime}) =e^{-\frac{\Gamma}{2}(t+t^{\prime})/\hbar}e^{\Gamma t_0/\hbar}\frac{n_d}{2}+ 2\Gamma\!\!\int\!\! \frac{d\epsilon}{2\pi} e^{i\left(\epsilon-\epsilon_d\right)\left(t-t'\right)/\hbar}g_{-}(\epsilon-\epsilon_d,t)g_{+}(\epsilon-\epsilon_d,t^{\prime}) f_\gamma(\epsilon),\\
&\Lambda^{(2)}_{\gamma}(t,t^{\prime}) = 2\int \!\frac{d\epsilon}{2\pi}  e^{i\left(\epsilon-\epsilon_d\right)\left(t-t'\right)/\hbar}g_{-}(\epsilon-\epsilon_d,t)f_\gamma\left(\epsilon\right),
\end{aligned}
\end{equation}
and
\begin{equation}
    g_{\pm}(\epsilon,t)=\frac{e^{-\left(\frac{\Gamma}{2}\pm i\epsilon\right)\frac{t-t_0}{2\hbar}}}{\frac{\Gamma}{2}\pm i\epsilon}\sinh{\left[\left(\frac{\Gamma}{2}\pm i\epsilon\right)\frac{t-t_0}{2\hbar}\right]},
\end{equation}
where $n_d$ is the initial population of the dot at the initial time $t_0$, and $\Gamma=\sum_\beta \Gamma_\beta$, is the total coupling rate.
The barred quantities in Eqs. \eqref{app_eq:fluctuation_Lambda}, correspond to the above expressions by simply substituting $n_d\rightarrow (1-n_d)$ and $f_{\gamma}\left(\epsilon \right)\rightarrow \left(1-f_{\gamma}\left(\epsilon \right)\right)$ with $n_d$ and $f_{\gamma}( \epsilon)=\{e^{(\epsilon-\mu_{\gamma})/ T_{\gamma}}+1\}^{-1}$ being respectively the initial population of the dot and the initial population of the reservoir $\gamma$ represented by the Fermi-Dirac function at the thermal equilibrium with temperature $T_\gamma$, and chemical potential $\mu_\gamma$.

Finally, substituting Eq. \eqref{app_eq:fluctuation_Lambda} into Eq. \eqref{app_eq:Activity}, we obtain the analytical form of the activity in the strong-coupling regime

\begin{equation}
\label{app_eq:Activity_analytical_strong_coupling}
\begin{aligned}
\mathcal{A}_\alpha(t) =&\sum_\beta \frac{\Gamma_\alpha \Gamma_\beta}{\hbar^2}\Re\int_{-t}^t d\tau  \Lambda_\alpha^{(0)}(t,t+\tau) \overline{\Lambda}_\beta^{(1)}(t+\tau,t)-\delta_{\beta\alpha}  \Lambda^{(2)}_\alpha(t,t+\tau) \overline{\Lambda}_\alpha^{(2)}(t+\tau,t)
+\{\Lambda \leftrightarrow\overline{\Lambda}\}.
\end{aligned}
\end{equation}

\subsection{Weak-Coupling Limit of the Activity for a SQD}
\label{app_weak_coupling_protocol_SQD}

To compute the activity in the weak-coupling regime, we follow the protocol detailed in Ref.~\cite{Blasi2024}. As explained therein, the weak-coupling expression of a two-point correlation function (such as the exchange-rate fluctuation) can be computed as follows:
\begin{equation}
\label{eq:QD-act-weak}
\mathcal{A}^{\wc}_\alpha(t,\tau)=\Gamma^2\lim_{\substack{\Gamma\to 0\\\Gamma t,\Gamma \tau\sim \text{const.}} }\frac{\mathcal{A}_\alpha(t,\tau)}{\Gamma^2}, 
\end{equation}
where the superscript $\wc$ stands for \textit{weak-coupling}.
Here, we consider the two-time activity $\mathcal{A}_\alpha(t,\tau)$, depending on both $t$ and $\tau$, and formally coinciding with the integrand of Eq.~(2) of the main text, i.e., 
\begin{equation}
\mathcal{A}_\alpha(t,\tau) \equiv \frac{\langle\langle \left\{\hat{V}_\alpha (t), \hat{V}_\alpha (t+\tau) \right\}\rangle\rangle}{2\hbar^2} \sim \mathcal{O}(\Gamma^2).
\end{equation}
The single-time activity defined in Eq.~(2) of the main text can then be obtained by integrating the two-time expression over $\tau$ from $-t$ to $t$. The weak-coupling protocol outlined in Eq.~\eqref{eq:QD-act-weak} involves the following steps: (i) divide the strong-coupling two-time activity by $\Gamma^2$; (ii) take the limit $\Gamma \to 0$, corresponding to the Born approximation, while keeping $\Gamma t$ and $\Gamma \tau$ constant—this reflects the Markovian approximation; and (iii) multiply the resulting expression by $\Gamma^2$ to isolate the correct leading-order contribution in the coupling.

Dividing $\mathcal{A}_\alpha(t,\tau)$ by $\Gamma^2$ and taking the limit $\Gamma \to 0$ (with $\Gamma t,\Gamma t'\sim \text{const.}$), corresponds to evaluating the weak-coupling limit of the $\Lambda$-functions defined in Eqs.~\eqref{Lambda_HE}, yielding:
\begin{equation}
    \label{Lambda_ME}
\begin{aligned}
    &\lim_{\substack{\Gamma\to 0\\\Gamma t,\Gamma t'\sim \text{const.}}}	\Lambda^{(0)}_{\gamma}(t,t^{\prime})= \frac{2\hbar}{\Gamma}f_{\gamma}(\epsilon_d)\delta (t-t'), \\
    &\lim_{\substack{\Gamma\to 0\\\Gamma t,\Gamma t'\sim \text{const.}}}	\Lambda^{(1)}_{\gamma}(t,t^{\prime})=e^{- \frac{\Gamma}{2} (t+t^{\prime})/\hbar}e^{\Gamma t_0/\hbar}n_d+ \left[\Theta (t^{\prime}-t)e^{- \frac{\Gamma}{2} \abs{t-t^{\prime}}/\hbar}-\Theta (t_0-t)e^{- \frac{\Gamma}{2} (t^{\prime}-t_0)/\hbar}e^{- \frac{\Gamma}{2} \abs{t-t_0}/\hbar}\right]f_{\gamma}(\epsilon_d),\\
    &\lim_{\substack{\Gamma\to 0\\\Gamma t,\Gamma t'\sim \text{const.}}}	\Lambda^{(2)}_{\gamma}(t,t^{\prime})= \frac{1}{2}\left[e^{- \frac{\Gamma}{2} \abs{t-t^{\prime}}/\hbar}+e^{- \frac{\Gamma}{2} (t+t^{\prime})/\hbar}e^{\Gamma t_0/\hbar}-e^{- \frac{\Gamma}{2} (t^{\prime}-t_0)/\hbar}e^{- \frac{\Gamma}{2} \abs{t-t_0}/\hbar}-e^{- \frac{\Gamma}{2} (t-t_0)}e^{- \frac{\Gamma}{2} \abs{t^{\prime}-t_0}/\hbar}\right]f_{\gamma}(\epsilon_d).
\end{aligned}
\end{equation}
The expressions in Eq.~\eqref{Lambda_ME} are obtained by shifting the integration variable as $\epsilon \to \epsilon + \epsilon_d$ and rescaling the energy according to
$\epsilon \to \Gamma \epsilon$. Taking the limit $\Gamma \to 0$ then allows one to factor out the Fermi functions evaluated at the dot energy $\epsilon_d$ and to perform the energy integrals analytically, yielding $\Lambda$-functions that depend only on the time variables. In this procedure, it is essential to keep the scaled times $\Gamma t$ and $\Gamma t'$ finite as $\Gamma \to 0$.
\\
Physically, the limit $\Gamma \to 0$ corresponds to the weak system--reservoir coupling underlying the Born approximation, while keeping $\Gamma t$ and $\Gamma t'$ constant selects the relevant long-time dynamics and reflects the Markovian approximation, in which reservoir correlations decay rapidly compared to the system timescale. Further details of this weak-coupling protocol can be found in Ref.~\cite{Blasi2024}.

By substituting these expressions into Eq.~\eqref{app_eq:Activity_analytical_strong_coupling} and performing the $\tau$-integral from $-t$ to $t$, it is possible to show analytically that the weak-coupling activity takes the form:
\begin{equation}
\label{WCactivity}
     \mathcal{A}^{\wc}_\alpha(t) = \sum_{\beta}\frac{\Gamma_{\alpha}\Gamma_{\beta}}{\hbar\Gamma}\left\{e^{-\Gamma t/\hbar}\left(f_{\beta}-n_d\right)\left[f_{\alpha}-(1-f_{\alpha})\right] +  \left[f_{\beta}\left(1-f_{\alpha}\right)+f_{\alpha}\left(1-f_{\beta}\right)\right]\right\},
\end{equation}
which exactly matches the result obtained via the master equation approach: \begin{eqnarray} \mathcal{A}^{\wc}_\alpha(t) \equiv \mathcal{A}^{\me}_\alpha(t). \end{eqnarray} Here, the superscript $\me$ refers to the master equation result, as given in Eq.~(90) of Ref.~\cite{Blasi2024}.
Interestingly, in the stationary limit, the weak-coupling activity in
Eq.~\eqref{WCactivity} has a clear physical interpretation in terms of steady-state
exchange processes between the reservoirs and the quantum dot, as noted in
Ref.~\cite{Blasi2024}. The overall prefactor, proportional to
$\Gamma_\alpha \Gamma_\beta / \Gamma$, represents the probability for a particle
exchange process involving reservoirs $\alpha$ and $\beta$. The terms inside the
brackets, such as $f_{\beta}(1-f_{\alpha})$ and $f_{\alpha}(1-f_{\beta})$,
correspond to occupation and deoccupation processes: they describe the probability
that an incoming particle from reservoir $\beta$ occupies the dot level, multiplied
by the probability that the corresponding state in reservoir $\alpha$ is available
(empty).

%%%%%%%%%%%%%%%%%%%%%%%%%%%%%%%%%%%%%%%%%%%%%%%%%%%%%%%%%%%%%%%%%%%%%%%%%%%%%%%%%%%%%%%

\section{Weak-coupling benchmark in the steady state: DQD}
\label{App:DQD}

In this section, we present the calculation of the activity for a double quantum dot (DQD) system attached to two reservoirs using the master equation (ME) formalism. We also compare this result with the Green's function approach in the weak-coupling regime.

\subsection{Activity for a DQD with ME}
Under weak-coupling between the system and reservoirs, the master equation (ME) describes the evolution of the reduced density operator $\rho$ via the Liouvillian superoperator $\mathcal{L}$. In Lindblad form, $\mathcal{L}$ includes both the unitary dynamics generated by the Hamiltonian  $\hat{H}_S=\sum_{n=1,2}\epsilon_{n} \hat{d}^{\dagger}_n\hat{d}_n + g\sum_{n\neq m}\hat{d}_n^{\dagger}\hat{d}_m$, where $\epsilon_1$, $\epsilon_2$ are the dot energies and $g$ the inter-dot tunneling amplitude, and the dissipative contributions: 
\begin{align}
\dot{\rho}(t) = \mathcal{L} \rho(t) = -\frac{i}{\hbar}[\hat{H}_S, \rho(t)] +\frac{1}{\hbar} \sum_{j\alpha} \big( \Gamma_{j\alpha}^+ \mathcal{D}[\hat{L}_{j\alpha}^\dagger] + \Gamma_{j\alpha}^- \mathcal{D}[\hat{L}_{j\alpha}] \big)\rho(t),
\end{align}
where $\hat{L}_{j\alpha}^\dagger$ and $\hat{L}_{j\alpha}$ are jump operators for excitation exchange between the system's energy $\epsilon_j$ and reservoir $\alpha$. The dissipators are defined as $\mathcal{D}[X]\rho \coloneqq X \rho X^\dagger - \frac{1}{2}(X^\dagger X \rho + \rho X^\dagger X)$, ensuring trace preservation.

For a double quantum dot (DQD) system coupled to two reservoirs $\alpha = L, R$, in the local ME approach, dot 1 (2) is coupled solely with the left (right) reservoir. Assuming energy-degenerate dots $\epsilon_1 = \epsilon_2 = \epsilon_d$, the rates and jump operators are:
\begin{align}
\label{rates_and_jump_op}
\Gamma_{j\alpha}^+ = \Gamma_\alpha f_\alpha(\epsilon_d), \quad \Gamma_{j\alpha}^- = \Gamma_\alpha \big(1 - f_\alpha(\epsilon_d)\big), \quad
\hat{L}_{jL} = \delta_{1j} \hat{\sigma}_-^{(j)}, \quad \hat{L}_{jR} = \delta_{2j} \hat{\sigma}_-^{(j)},
\end{align}
with $\hat{\sigma}_\pm^{(1)} \equiv \hat{\sigma}_\pm \otimes \mathds{1}$ and $\hat{\sigma}_\pm^{(2)} \equiv \mathds{1} \otimes \hat{\sigma}_\pm$.
Following Ref.~\cite{Blasi2024}, the ME stationary activity $\mathcal{A}_{\text{ME},\alpha}^{\text{ss}}$ in the reservoir $\alpha$, is computed as the sum of the incoming and outgoing contributions from all jump processes in the steady state:
\begin{align}
\mathcal{A}_{\text{ME}}^{\text{ss}} = \frac{1}{\hbar}\Tr\left\{\big(\mathcal{L}_\alpha^+ + \mathcal{L}_\alpha^-\big)\rho_{\text{ss}}\right\},
\end{align}
where $\mathcal{L}_\alpha^+$ and $\mathcal{L}_\alpha^-$ are jump superoperators defined as:
\begin{align}
\mathcal{L}^+_{j\alpha}\rho = \Gamma_{j\alpha}^+ \hat{L}_{j\alpha}^\dagger \rho \hat{L}_{j\alpha}, \quad
\mathcal{L}^-_{j\alpha}\rho = \Gamma_{j\alpha}^- \hat{L}_{j\alpha} \rho \hat{L}_{j\alpha}^\dagger.
\end{align}
The superoperator $\mathcal{L}^+_{j\alpha}$ describes the process of a particle tunneling into the quantum system from reservoir $\alpha$ to the energy level $\epsilon_j$ at a rate $\Gamma_{j\alpha}^+$. Conversely, $\mathcal{L}^-_{j\alpha}$ captures the tunneling of a particle out of the quantum system from energy level $\epsilon_j$ to reservoir $\alpha$, occurring at a rate $\Gamma_{j\alpha}^-$. 
Using the rates and jump operators defined in Eq.~\eqref{rates_and_jump_op}, the stationary activity in the left reservoir $\alpha=L$ can be expressed explicitly as:

\begin{align}
\mathcal{A}_{L}^{\me, ss} = 
2\frac{\Gamma_L^2}{\hbar \Gamma}\frac{4g^2 + \Gamma_R \Gamma}{4g^2 + \Gamma_L \Gamma_R} f_L(1 - f_L) 
+ 4\frac{g^2 \Gamma_L \Gamma_R}{\hbar \Gamma(4g^2 + \Gamma_L \Gamma_R)}\big[f_L(1 - f_R) + f_R(1 - f_L)\big].
\end{align}
Similar expressions can be obtained for the right reservoir and the total activity.

%%%%%%%%%%%%%%%%%%%%%%%%%

\subsection{From Green's functions to the weak-coupling limit}

\noindent The retarded and advanced Green's functions for a DQD system in a series configuration are given by~\cite{Covito2018}:
\begin{equation}
    \mathbf{G}^r(\epsilon) = \mathbf{G}^a(\epsilon)^\dagger = \frac{1}{\Omega}
    \begin{pmatrix}
        \epsilon - \epsilon_2 + i \frac{\mathbf{\Gamma}_{22}}{2} & -g + i \frac{\mathbf{\Gamma}_{21}}{2} \\[8pt]
        -g + i \frac{\mathbf{\Gamma}_{12}}{2} & \epsilon - \epsilon_1 + i \frac{\mathbf{\Gamma}_{11}}{2}
    \end{pmatrix},
\end{equation}
where 
\begin{equation}
    \Omega = \left(\epsilon - \epsilon_1 + i \frac{\mathbf{\Gamma}_{11}}{2}\right)
    \left(\epsilon - \epsilon_2 + i \frac{\mathbf{\Gamma}_{22}}{2}\right) 
    - \left(-g + i \frac{\mathbf{\Gamma}_{12}}{2}\right)\left(-g + i \frac{\mathbf{\Gamma}_{21}}{2}\right),
\end{equation}
and the coupling matrices are defined as:
\begin{equation}
    \mathbf{\Gamma}_L = 
    \begin{pmatrix}
        \Gamma_L & 0 \\
        0 & 0
    \end{pmatrix}, \quad
    \mathbf{\Gamma}_R = 
    \begin{pmatrix}
        0 & 0 \\
        0 & \Gamma_R
    \end{pmatrix}, \quad
    \mathbf{\Gamma} = \mathbf{\Gamma}_L + \mathbf{\Gamma}_R.
\end{equation}
Using these relations, we can compute the T-matrices introduced in Eq.~\eqref{eq:T-matrices} \begin{eqnarray}
\mathbf{T}_{\alpha\beta}(\epsilon) = \mathbf{\Gamma}_\alpha(\epsilon)\mathbf{G}^r(\epsilon) \mathbf{\Gamma}_\beta(\epsilon)  \mathbf{G}^a(\epsilon),
\end{eqnarray} 
which are used to calculate the activity as given in Eq.~\eqref{eq: activityGreenmultiterminal}. 
As an example, the transmission function for degenerate dots (\(\epsilon_1 = \epsilon_2 = \epsilon_d\)), takes the following form:
\begin{equation}
    \mathcal{T}_{RL}(\epsilon) = \Tr[\mathbf{T}_{RL}(\epsilon)] = \Tr[\mathbf{T}_{LR}(\epsilon)] 
    = \frac{g^2 \Gamma_L \Gamma_R}{\big(g^2 + \frac{\Gamma_L \Gamma_R}{4} - (\epsilon - \epsilon_d)^2\big)^2 + \frac{\Gamma^2}{4} (\epsilon - \epsilon_d)^2}.
\end{equation}
The activity at the left lead ($\alpha = L$) can be expressed as 
\begin{align} 
\label{eq:Ass_dqd}
    \mathcal{A}^{ss}_L =\frac{1}{\hbar} \int \frac{d\epsilon}{2\pi} 
    & \frac{4 \Gamma_L^2 \big( -g^2 + \frac{\Gamma_R^2}{4} + (\epsilon - \epsilon_d)^2 \big)^2 (\epsilon - \epsilon_d)^2}{
        \left[\big(g^2 + \frac{\Gamma_L \Gamma_R}{4} - (\epsilon - \epsilon_d)^2\big)^2 
        + \frac{\Gamma^2}{4} (\epsilon - \epsilon_d)^2
    \right]^2} F_{LL}(\epsilon) \nonumber \\
    & + \frac{g^2 \Gamma_L \Gamma_R}{
        \big(g^2 + \frac{\Gamma_L \Gamma_R}{4} - (\epsilon - \epsilon_d)^2\big)^2 
        + \frac{\Gamma^2}{4} (\epsilon - \epsilon_d)^2
    } \big[ F_{LR}(\epsilon) + F_{RL}(\epsilon) \big].
\end{align}
Following Ref.~\cite{Blasi2024}, to take the weak-coupling (WC) limit of the steady-state activity, we divide the above expression by $\Gamma$, take the limit $\Gamma \to 0$, and then multiply back by $\Gamma$ to retain the correct coupling order:
\begin{align}
\label{eq:Ass_dqd_wc}
    \mathcal{A}^{\wc, ss}_{L} = \Gamma \lim_{\Gamma \to 0} \frac{\mathcal{A}^{ss}_L}{\Gamma}.
\end{align}
This differs from the case of the two-time activity in Eq.~\eqref{eq:QD-act-weak}, since the integration over $\tau$ (corresponding to the zero frequency component) makes the steady-state activity scale as $\mathcal{O}(\Gamma)$, in contrast to the $\mathcal{O}(\Gamma^2)$ scaling of the two-time activity.
By inserting Eq.~\eqref{eq:Ass_dqd} into Eq.~\eqref{eq:Ass_dqd_wc}, shifting the energies by $\epsilon_d$, and performing the change of variable $\epsilon \to \Gamma \omega$ (with $\omega$ an adimensional variable), we obtain 
\begin{align}
    \mathcal{A}^{\wc, ss}_{L} =  \Gamma \lim_{\Gamma \to 0} \frac{1}{\hbar}\int \frac{d\omega}{2\pi}& \frac{\Gamma_L^2}{\Gamma^2} 
    \frac{4 \omega^2 \big( -\frac{g^2}{\Gamma^2} + \frac{\Gamma_R^2}{\Gamma^2} + \omega^2 \big)^2}{
        \left[\big( \frac{g^2}{\Gamma^2} + \frac{\Gamma_L \Gamma_R}{4 \Gamma^2} - \omega^2 \big)^2 
        + \frac{\omega^2}{4}
    \right]^2} F_{LL}(\Gamma \omega + \epsilon_d) \nonumber \\
    & + \frac{\Gamma_L \Gamma_R}{\Gamma^2} 
    \frac{\big( \frac{g}{\Gamma} \big)^2}{
        \big( \frac{g^2}{\Gamma^2} + \frac{\Gamma_L \Gamma_R}{4 \Gamma^2} - \omega^2 \big)^2 
        + \frac{\omega^2}{4}
    } \big[ F_{LR}(\Gamma \omega + \epsilon_d) + F_{RL}(\Gamma \omega + \epsilon_d) \big].
\end{align}
where we recall the definition on the statistical factors
$F_{\alpha\beta}(\epsilon)=f_\alpha(\epsilon)\bigl[1-f_\beta(\epsilon)\bigr]$, expressed in terms of the Fermi--Dirac distributions.
In the weak-coupling limit $\Gamma \to 0$, assuming $g/\Gamma$ remains constant (as required in the local master equation regime), the Fermi functions can be evaluated at $\epsilon_d$ and factored out of the integral. The resulting expression can be explicitly integrated to recover the result obtained from the Master Equation:
\begin{equation}
    \mathcal{A}^{\wc, ss}_{L} = \mathcal{A}^{\me, ss}_{L}.
\end{equation}
The same procedure applies to the activity at the right lead and, consequently, to the total activity.

%%%%%%%%%%%%%%%%%%%%%%%%%%%%%%%%%%%%%%%%%%%%%%%%%%%%%%%%%%%%%%%%%%%%%%%%%%%%%%%%%%%%%%%%%%%%%%%%

\section{Multi-dot steady state Activity with Green's functions}
\label{app_stationary_state}
In the stationary regime, the steady state activity for a multi-dot system (with $i,j=1,\cdots, D$ labeling the dots), can be obtained as the limit of $t\to \infty$ of Eq. \eqref{app_eq:Activity}, which can be expressed in the following fashion
\begin{align}
\label{app_eq:Activity_stationary}
\mathcal{A}^{ss}_\alpha = \lim_{t \to \infty} \mathcal{A}_\alpha(t) = \sum_{ij}\sum_{kk'} \int d\tau~ & \left\{ 
    t^*_{i k\alpha} t^*_{j k'\alpha} \calg_{j,k\alpha}^<(\tau) \calg_{i,k'\alpha}^>(-\tau) +   
    t^*_{i k\alpha} t_{j k'\alpha} \calg_{k'\alpha,k\alpha}^<(\tau) \calg_{i,j}^>(-\tau)  
    \right. \nonumber \\
    & \left. +  t_{i k\alpha} t^*_{j k'\alpha} \calg_{j,i}^<(\tau) \calg_{k\alpha,k'\alpha}^>(-\tau) +   
    t_{i k\alpha} t_{j k'\alpha} \calg_{k\alpha,i}^<(\tau) \calg_{k'\alpha,j}^>(-\tau) 
\right\}.
\end{align}
Here, we omitted the $ ``\Re"$ since the integral is real, and introduced the standard lesser and greater Green's functions. 
In the stationary regime, where time translational invariance applies, these functions depend solely on time difference $t-t'\equiv\tau$ and are defined as follows:
\begin{align}
\label{Green_functions}
\calg^{<}_{j,k\alpha}(t-t')&=\frac{i}{\hbar}\expval{\hat{c}_{k\alpha}^{\dagger}(t')\hat{d}_j(t)},& \calg^{>}_{k'\alpha, j}(t-t')&=-\frac{i}{\hbar}\expval{\hat{c}_{k'\alpha}(t) \hat{d}_j^{\dagger}(t^{\prime})},\nonumber\\
 \calg^{<}_{k\alpha,i}(t-t')&=\frac{i}{\hbar}\expval{ \hat{d}_i^{\dagger}(t')\hat{c}_{k\alpha}(t)},&\calg^{>}_{i,\alpha k'}(t-t')&=-\frac{i}{\hbar}\expval{\hat{d}_i(t)\hat{c}_{k'\alpha}^{\dagger}(t^{\prime})},\nonumber\\
\calg^{<}_{k^{\prime}\alpha,k\alpha}(t-t')&=\frac{i}{\hbar}\expval{\hat{c}_{k\alpha}^{\dagger}(t')\hat{c}_{k^{\prime}\alpha}(t)},& \calg^{>}_{k\alpha,k^{\prime}\alpha}(t-t')&=-\frac{i}{\hbar}\expval{\hat{c}_{k\alpha}(t)c^{\dagger}_{k^{\prime}\alpha}(t^{\prime})},\nonumber\\
\calg^{<}_{j,i}(t-t')&=\frac{i}{\hbar}\expval{ \hat{d}_i^{\dagger}(t')\hat{d}_j(t)},&\calg^{>}_{i,j}(t-t')&=-\frac{i}{\hbar}\expval{\hat{d}_i(t) \hat{d}_j^{\dagger}(t^{\prime})}.
\end{align}
By noticing that Eq. \eqref{app_eq:Activity_stationary}, corresponds to the zero frequency component of products of Green's functions, using the convolution theorem, we can express it in the Fourier space as
\begin{align}
\label{app_eq:Activity_stationary_Fourier}
\mathcal{A}^{ss}_\alpha =\frac{1}{\hbar} \sum_{ij}\sum_{kk'} \int \frac{d\epsilon}{2\pi}~ & \left\{ 
    t^*_{i k\alpha} t^*_{j k'\alpha} \calg_{j,k\alpha}^<(\epsilon) \calg_{i,k'\alpha}^>(\epsilon) +   
    t^*_{i k\alpha} t_{j k'\alpha} \calg_{k'\alpha,k\alpha}^<(\epsilon) \calg_{i,j}^>(\epsilon)  
    \right. \nonumber \\
    & \left. +  t_{i k\alpha} t^*_{j k'\alpha} \calg_{j,i}^<(\epsilon) \calg_{k\alpha,k'\alpha}^>(\epsilon) +   
    t_{i k\alpha} t_{j k'\alpha} \calg_{k\alpha,i}^<(\epsilon) \calg_{k'\alpha,j}^>(\epsilon) 
\right\}.
\end{align}
In the Keldysh formalism, since the Hamiltonian describing the leads is noninteracting, one has the Dyson equations~\cite{Meir1992}
\begin{align}
    \calg_{i,k\alpha}^{\lessgtr}(\epsilon)=&\sum_{n}\Big\{ \calg_{i,n}^{\lessgtr}(\epsilon)t_{n k\alpha}g_{k\alpha}^a(\epsilon)+\calg_{i,m}^{r}(\epsilon)t_{n k\alpha}g_{k\alpha}^\lessgtr(\epsilon)\Big\}, \nonumber \\
    \calg_{k\alpha,i}^{\lessgtr}(\epsilon)=&\sum_{n}\Big\{ g_{k\alpha}^\lessgtr(\epsilon)t_{n k\alpha}^*\calg_{n,i}^{a}(\epsilon)+g_{k\alpha}^r (\epsilon)t_{n k\alpha}^*\calg_{m,i}^{\lessgtr}(\epsilon)\Big\}, \nonumber\\
    \calg_{k\alpha,k'\alpha}^{\lessgtr}(\epsilon)=&\sum_{n m}\Big\{ g_{k\alpha}^\lessgtr(\epsilon) t_{n k\alpha}^* \calg_{n,m}^{a}(\epsilon) t_{m k'\alpha} g_{k'\alpha}^a(\epsilon)+
    g_{k\alpha}^r(\epsilon) t_{n k\alpha}^* \calg_{n,m}^{\lessgtr}(\epsilon) t_{m k'\alpha} g_{k'\alpha}^a(\epsilon) \nonumber \\
    &+g_{k\alpha}^r(\epsilon) t_{n k\alpha}^* \calg_{n,m}^{r}(\epsilon) t_{m k'\alpha} g_{k'\alpha}^{\lessgtr}(\epsilon)
    +g_{k\alpha}^{\lessgtr}(\epsilon)
    \Big\}.
\end{align}
expressed in terms of the retarded $\calg_{i,j}^{r}$ and  advanced $\calg_{i,j}^{a}$ multi-dot Green's functions, and the unperturbed Green's functions $g_{k\alpha,d}^{r,a}(\epsilon)=\mp \pi i\delta(\epsilon-\epsilon_{k\alpha})$, $g_{k\alpha,d}^{<}(\epsilon)=2\pi i f_{\alpha}(\epsilon)\delta(\epsilon-\epsilon_{k\alpha})$ and $g_{k\alpha,d}^{>}(\epsilon)=-2 \pi i (1-f_{\alpha}(\epsilon))\delta(\epsilon-\epsilon_{k\alpha})$.

By substituting the above relations in the expression of the stationary activity, one can rewrite the four terms in  Eq. \eqref{app_eq:Activity_stationary} in the following compact fashion 

\begin{align}\label{noise4terms}
\sum_{ij}\sum_{k k'} t^*_{i k\alpha} t^*_{j k'\alpha} \calg_{j,k\alpha}^<(\epsilon) \calg_{i,k'\alpha}^>(\epsilon)
=&
\Tr\Big\{\left[\mathbf{G^<}(\epsilon)\mathbf{\Sigma_{\alpha}^a}(\epsilon)+ \mathbf{G^r}(\epsilon)\mathbf{\Sigma_{\alpha}^<}(\epsilon) \right]\times
\left[ \mathbf{G^>}(\epsilon)\mathbf{\Sigma_{\alpha}^a}(\epsilon)+\mathbf{G^r}(\epsilon)\mathbf{\Sigma_{\alpha}^>}(\epsilon) \right]\Big\} \nonumber,
\\
\sum_{ij}\sum_{k k'} t^*_{i k\alpha} t_{j k'\alpha} \calg_{k'\alpha,k\alpha}^<(\epsilon) \calg_{i,j}^>(\epsilon)
=&
\Tr \Big\{\mathbf{\Sigma_{\alpha}^<}(\epsilon)\mathbf{G^>}(\epsilon) + \mathbf{\Sigma_{\alpha}^r}(\epsilon)\left[\mathbf{G^r}(\epsilon)\mathbf{\Sigma_{\alpha}^<}(\epsilon) 
+ \mathbf{G^<}(\epsilon)\mathbf{\Sigma_{\alpha}^a}(\epsilon)\right]\mathbf{G^>}(\epsilon)
\nonumber \\
& + \mathbf{\Sigma_{\alpha}^<}(\epsilon)\mathbf{G^a}(\epsilon)\mathbf{\Sigma_{\alpha}^a}(\epsilon)\mathbf{G^>}(\epsilon)\Big\} \nonumber,
\\
\sum_{ij}\sum_{kk'} t_{i k\alpha} t^*_{j k'\alpha} \calg_{j,i}^<(\epsilon) \calg_{k\alpha,k'\alpha}^>(\epsilon)
=&
\Tr \Big\{\mathbf{G^<}(\epsilon)\mathbf{\Sigma_{\alpha}^>}(\epsilon) + \mathbf{G^<}\mathbf{\Sigma_{\alpha}^r}(\epsilon)\left[\mathbf{G^r}(\epsilon)\mathbf{\Sigma_{\alpha}^>}(\epsilon) 
+ \mathbf{G^>}(\epsilon)\mathbf{\Sigma_{\alpha}^a}(\epsilon)\right] 
 \nonumber \\
& + \mathbf{G^<}\mathbf{\Sigma_{\alpha}^>}(\epsilon)\mathbf{G^a}(\epsilon)\mathbf{\Sigma_{\alpha }^a}(\epsilon)\Big\}\nonumber,
\\
\sum_{i j}\sum_{k k'} t_{i k\alpha} t_{j k'\alpha} \calg_{k\alpha,i}^<(\epsilon) \calg_{k'\alpha,j}^>(\epsilon)
=&
\Tr \Big\{\left[ \mathbf{\mathbf{\Sigma_{\alpha}^<}(\epsilon)\mathbf{G^a}(\epsilon)+\Sigma_{\alpha}^r}(\epsilon)\mathbf{G^<}(\epsilon) \right]\times
\left[ \mathbf{\mathbf{\Sigma_{\alpha}^>}(\epsilon)\mathbf{G^a}(\epsilon)+\Sigma_{\alpha}^r}(\epsilon)\mathbf{G^>}(\epsilon) \right]\Big\},
\end{align}
where we introduced the multi-dot Green's functions matrix $\left[\mathbf{G^{a,r,\lessgtr}}\right]_{ij}=\calg_{i,j}^{a,r,\lessgtr}$, and self-energies matrices in the reservoir $\alpha$
\begin{equation}
\label{sigmaanalytical}
\begin{aligned}
    &\left[\mathbf{\Sigma_{\alpha}^r}(\epsilon)\right]_{ij} = \sum_k t_{i k\alpha}t_{j k\alpha}^* g_{k\alpha}^r(\epsilon)  = -i\frac{\left[\mathbf{\Gamma_{\alpha}}(\epsilon) \right]_{ij}}{2},\quad  &&\left[\mathbf{\Sigma_{\alpha}^a}(\epsilon)\right]_{ij} = \sum_k t_{i k\alpha}t_{j k\alpha}^* g_{k\alpha}^a(\epsilon)  = i\frac{\left[\mathbf{\Gamma_{\alpha}}(\epsilon) \right]_{ij}}{2},\\
    &\left[\mathbf{\Sigma_{\alpha}^<}(\epsilon)\right]_{ij} = \sum_k t_{i k\alpha}t_{j k\alpha}^* g_{k\alpha}^<(\epsilon)  = if_{\alpha}(\epsilon)\left[\mathbf{\Gamma_{\alpha}}(\epsilon) \right]_{ij},\quad
&&\left[\mathbf{\Sigma_{\alpha}^>}(\epsilon)\right]_{ij} = \sum_k t_{i k\alpha}t_{j k\alpha}^* g_{k\alpha}^>(\epsilon)  = -i\left(1- f_{\alpha}(\epsilon)\right)\left[\mathbf{\Gamma_{\alpha}}(\epsilon) \right]_{ij},\\
\end{aligned}
\end{equation}
with the dot-reservoir coupling matrix elements $\left[\mathbf{\Gamma_{\alpha}(\epsilon) }\right]_{ij} = 2\pi\sum_k t_{i k\alpha} t_{j k\alpha}^* \delta(\epsilon-\epsilon_{\alpha k})$. 
The function $f_\alpha(\epsilon)$ corresponds to the Fermi distribution for fermionic statistics. 
Finally, making use of the relations  
\begin{align}
 &\mathbf{G^{\lessgtr}}(\epsilon)= \sum_\beta\mathbf{G^{r}}(\epsilon) \mathbf{\Sigma_{\beta}^\lessgtr}(\epsilon)\mathbf{G^{a}}(\epsilon),\\
&i\left[\mathbf{G^{a}}(\epsilon)-\mathbf{G^{r}}(\epsilon)\right]=-\sum_\beta\mathbf{G^{r}}(\epsilon) \mathbf{\Gamma_{\beta}}(\epsilon)\mathbf{G^{a}}(\epsilon),
\end{align}
we can write the steady state activity in the reservoir $\alpha=1,\cdots,N$ (with $N$ the total number of terminals) for a multi-dot quantum system 
\begin{align}
\label{eq: activityGreenmultiterminal}
\mathcal{A}^{ss}_\alpha = \frac{1}{\hbar}\int \frac{d\epsilon}{2\pi}~ & \Tr{4\mathbf{T}_{\alpha\alpha}(\epsilon) - \left(\sum_{\beta} \mathbf{T}_{\alpha\beta}(\epsilon)\right)^2} F_{\alpha\alpha}(\epsilon) + \sum_{\beta\neq\alpha}\Tr{\mathbf{T}_{\alpha\beta}(\epsilon)}\left[F_{\alpha\beta}(\epsilon) + F_{\beta\alpha}(\epsilon)\right],
\end{align}
where $F_{\alpha\beta}(\epsilon) = f_{\alpha}(\epsilon)(1-f_\beta(\epsilon))$. 
The matrices are defined as:
\begin{eqnarray}
\label{eq:T-matrices}
\mathbf{T}_{\alpha\beta}(\epsilon) = \mathbf{\Gamma}_\alpha(\epsilon)\mathbf{G}^r(\epsilon) \mathbf{\Gamma}_\beta(\epsilon)  \mathbf{G}^a(\epsilon).
\end{eqnarray} 
It is important to note that $\mathcal{T}_{\alpha\beta}(\epsilon) = \Tr{\mathbf{T}_{\alpha\beta}(\epsilon)}$ corresponds to the standard transmission function from reservoir $\alpha$ to reservoir $\beta$ commonly defined in transport. 
Summing over the lead's index $\alpha$, and using that $F_{\alpha\beta}(\epsilon)+F_{\beta\alpha}(\epsilon)=(f_\alpha(\epsilon)-f_\beta(\epsilon))^2+F_{\alpha\alpha}(\epsilon)+F_{\beta\beta}(\epsilon)$, we can write the total activity in the following way 
\begin{align}
\label{eq: TotalactivityGreenmultiterminal}
\mathcal{A}^{ss} = \frac{1}{\hbar}\sum_{\alpha}\int \frac{d\epsilon}{2\pi}~ & \Tr{2\mathbf{T}_{\alpha\alpha}(\epsilon)+\sum_{\beta}\left(\mathbf{T}_{\alpha\beta}(\epsilon)+\mathbf{T}_{\beta\alpha}(\epsilon)\right) - \left(\sum_{\beta} \mathbf{T}_{\alpha\beta}\right)^2} F_{\alpha\alpha} + \sum_{\beta\neq\alpha}\Tr{\mathbf{T}_{\alpha\beta}(\epsilon)}\left(f_\alpha(\epsilon)-f_\beta(\epsilon) \right)^2.
\end{align}

A parallel derivation based on the Green's function approach can be carried out for the zero-frequency current autocorrelation, as detailed in Ref.~\cite{Blasi2024}.
By following steps analogous to those used to derive Eq.~\eqref{eq: activityGreenmultiterminal}, the zero-frequency noise in the steady state can be expressed as:
\begin{align}
\label{eq: NoiseGreenmultiterminal}
S_{\alpha\alpha} = \frac{1}{\hbar}\int \frac{d\epsilon}{2\pi}&~ \Tr{\left(\sum_{\beta\neq\alpha}\mathbf{T}_{\beta\alpha}(\epsilon)\right)^2} F_{\alpha\alpha}(\epsilon)  + 
\sum_{\beta\neq\alpha}\Tr{\mathbf{T}_{\beta\alpha}(\epsilon)\left(\mathbf{1} - \sum_{\beta\neq\alpha}\mathbf{T}_{\beta\alpha}(\epsilon)\right)}\left[F_{\alpha\beta}(\epsilon) + F_{\beta\alpha}(\epsilon)\right]\nonumber\\
&+ \sum_{\beta,\gamma\neq \alpha}\Tr{\mathbf{T}_{\beta\alpha}(\epsilon)\mathbf{T}_{\gamma\alpha}(\epsilon)} F_{\beta\gamma}(\epsilon).
\end{align}

%%%%%%%%%%%%%%%%%%%%%%%%%%%%%%%%%%%%%%%%%%%%%%%%%%%%%%%%%%%%%%%%%%%%%%%%%%%%%%%%%%%%%%%%%%%%%%%%%%%%%%%%%%%%%%%%%%%%%%%%%%%%%%%%%%%%%%%%%%%%%%%%%%%%%%%%%%%
\section{Activity in terms of the Scattering Matrix Amplitudes}\label{app_Green_in_scatt}

The Fisher-Lee relation~\cite{Fisher1981,Lopez2004,Sanchez2025} connects the retarded Green's function \(\mathcal{G}^r(\epsilon)\) to the scattering matrix \(s(\epsilon)\) of a mesoscopic quantum system in a multi-terminal configuration:
\begin{equation}
    \label{eq:Fisher_Lee_multiterminal}
    s_{\alpha\beta}(\epsilon)=\delta_{\alpha\beta}-i\sqrt{\Gamma_\alpha\Gamma_\beta}~\mathcal{G}^r_{\alpha\beta}(\epsilon).
\end{equation}
This relation holds under the condition \(\left[\Gamma_\alpha\right]_{ij}=\delta_{ij}\delta_{i\alpha}\Gamma_\alpha\), meaning that each terminal couples to a single quantum dot only~\cite{Boumrar2020}.
Substituting Eq.~\eqref{eq:Fisher_Lee_multiterminal} into Eq.~\eqref{eq: activityGreenmultiterminal}, we can express the stationary activity as the sum of an auto- and a cross-contribution, given in the main text by Eqs.~(8) and~(9):
\begin{equation}
\label{eq_app:auto_cross}
\mathcal{A}^{ss}_\alpha=\mathcal{A}_{\alpha}^{auto}+\mathcal{A}_{\alpha}^{cross}~~\text{with} ~~ \left\{
\begin{aligned}
\mathcal{A}_{\alpha}^{auto} &=  \frac{2}{\hbar}\int \frac{d\epsilon}{2\pi} \mathcal{R}_{\alpha\alpha}(\epsilon)\left(1-\cos{(\phi_\alpha)}\right) F_{\alpha\alpha}(\epsilon), \\
\mathcal{A}_{\alpha}^{cross} &= \frac{1}{\hbar} \sum_{\beta\neq \alpha}\int \frac{d\epsilon}{2\pi} \mathcal{T}_{\alpha\beta}(\epsilon) \left(F_{\alpha\beta}(\epsilon)+F_{\beta\alpha}(\epsilon)\right).
\end{aligned}
\right.
\end{equation}
Here, we used the expression for the reflection amplitude at lead \(\alpha\), \(s_{\alpha\alpha}\equiv r_{\alpha\alpha}=\sqrt{\mathcal{R}_{\alpha\alpha}}e^{i\phi_{\alpha}/2}\), where \(\mathcal{R}_{\alpha\alpha}\) denotes the reflection probability for a particle to remain in the same lead. Similarly, \(|{s_{\alpha\beta}}|^2\equiv \mathcal{T}_{\alpha\beta}\) gives the transmission probability from lead \(\beta\) to lead \(\alpha\).

Alternatively, by using the identity \(F_{\alpha\beta}(\epsilon)+F_{\beta\alpha}(\epsilon)=(f_\alpha(\epsilon)-f_\beta(\epsilon))^2+F_{\alpha\alpha}(\epsilon)+F_{\beta\beta}(\epsilon)\), the stationary activity can be decomposed into thermal and shot contributions, as shown in Eqs.~(6) and ~(7) of the main text:
\begin{equation}
\label{eq_app:th_sh}
\mathcal{A}^{ss}_\alpha=\mathcal{A}_{\alpha}^{th}+\mathcal{A}_{\alpha}^{sh}~~\text{with} ~~ \left\{
\begin{aligned}
\mathcal{A}_{\alpha}^{th} &= \frac{1}{\hbar}\int \frac{d\epsilon}{2\pi} \left[1+\mathcal{R}_{\alpha\alpha}(\epsilon)\left(1-2\cos{(\phi_\alpha)}\right)\right] F_{\alpha\alpha}(\epsilon)+\sum_{\beta\neq \alpha} \mathcal{T}_{\alpha\beta}(\epsilon)F_{\beta\beta}(\epsilon),\\
\mathcal{A}_{\alpha}^{sh} &=  \frac{1}{\hbar} \sum_{\beta\neq \alpha}\int \frac{d\epsilon}{2\pi} \mathcal{T}_{\alpha\beta}(\epsilon) \left(f_\alpha(\epsilon)-f_\beta(\epsilon)\right)^2.
\end{aligned}\right.
\end{equation}

Applying the same procedure to the autocorrelation noise, we substitute Eq.~\eqref{eq:Fisher_Lee_multiterminal} into Eq.~\eqref{eq: NoiseGreenmultiterminal}. This allows us to separate the steady-state current noise into two distinct contributions, commonly referred to as the classical and quantum components. This decomposition is consistent with the results obtained in Refs.~\cite{Palmqvist2025,Splettstoesser2024} using the Landauer-Büttiker formalism~\cite{Blanter2000}:
\begin{equation}
\label{eq_app:Noise_cl_qu}
S_{\alpha\alpha}=S_{\alpha\alpha}^{cl}-S_{\alpha\alpha}^{qu}~~\text{with} ~~ \left\{
\begin{aligned}
S_{\alpha\alpha}^{cl} &= \frac{1}{\hbar}\sum_{\beta\neq \alpha} \int \frac{d\epsilon}{2\pi}  \mathcal{T}_{\alpha\beta}(\epsilon)\left(F_{\alpha\beta}(\epsilon)+F_{\beta\alpha}(\epsilon)\right),\\
S_{\alpha\alpha}^{qu} &=  \frac{1}{\hbar} \int \frac{d\epsilon}{2\pi} \left(\sum_{\beta\neq \alpha} \mathcal{T}_{\alpha\beta}(\epsilon) \left(f_\alpha(\epsilon)-f_\beta(\epsilon)\right)\right)^2.
\end{aligned}\right.
\end{equation}
The terminology classical and quantum simply indicates that the classical term is linear in the transmission function and captures contributions from uncorrelated single-particle processes, while the quantum term is quadratic and encodes the effect of quantum correlations between particles.

\end{document}